\documentclass[%
 aip,
 sd,%
 amsmath,amssymb,
preprint,%
]{revtex4-1}

\usepackage{graphicx}
\usepackage{dcolumn}
\usepackage{bm}

\usepackage{graphicx,epic,amscd,amssymb,amsmath,multirow,color}
\usepackage[all]{xy}
\usepackage{subfigure}

\newcommand{\DIS}{\displaystyle}
\newcommand{\qed}{\hfill $\Box$}
\def\C{{\mathbb C}}
\def\Z{{\mathbb Z}}
\def\ZP{{\mathbb {ZP}}}
\def\P{{\mathbb P}}
\def\Q{{\mathbb Q}}
\def\T{{\mathbb T}}

\def\R{{\mathbb R}}

\def\bb{\mbox{\boldmath $b$}}
\def\bx{\mbox{\boldmath $x$}}
\def\by{\mbox{\boldmath $y$}}
\def\bz{\mbox{\boldmath $z$}}

\def\blambda{\mbox{\boldmath $\lambda$}}
\def\bSigma{\mbox{\boldmath $\Sigma$}}

\newcommand{\ctext}[1]{\raise0.2ex\hbox{\textcircled{\scriptsize{#1}}}}

\newtheorem{thm}{Theorem}
\newtheorem{corollary}{Corollary}
\newtheorem{remark}{Remark}

\newtheorem{lemma}{Lemma}
\newtheorem{proposition}{Proposition}

\begin{document}

\preprint{
}

\title[]{Periodicity, linearizability and integrability in seed mutations of type $\boldsymbol{A^{(1)}_N}$}

\author{Atsushi Nobe}
\email{nobe@faculty.chiba-u.jp}
\affiliation{ 
Faculty of Education, Chiba University,\\ 1-33 Yayoi-cho Inage-ku, Chiba 263-8522, Japan
}%
\author{Junta Matsukidaira}
\email{junta@rins.ryukoku.ac.jp}
\affiliation{ 
Department of Applied Mathematics and Informatics, Ryukoku University,\\ 1-5 Yokotani, Seta Oe-cho, Otsu, Shiga 520-2194, Japan
}%

\date{\today}

\begin{abstract}
In the network of seed mutations arising from a certain initial seed, an appropriate path emanating from the initial seed  is intendedly chosen, noticing periodicity of the exchange matrices in the path each of which is assigned to the generalized Cartan matrix of type $A^{(1)}_N$.
Then dynamical property of the seed mutations along the path, which is referred to as of type $A^{(1)}_N$, is intensively investigated.
The coefficients assigned to the path form certain $N$ monomials that posses periodicity with period $N$ under the seed mutations and enable to obtain the general terms of the coefficients.
The cluster variables assigned to the path of type $A^{(1)}_N$ also form certain $N$ Laurent polynomials possessing the same periodicity as the monomials generated  by the coefficients.
These Laurent polynomials lead to sufficiently number of conserved quantities of the dynamical system derived from the cluster mutations along the path.
Furthermore, by virtue of the Laurent polynomials with periodicity, the dynamical system is non-autonomously linearized and its general solution is concretely constructed.
Thus the seed mutations along the path of type $A^{(1)}_N$ exhibit discrete integrability.
%
\end{abstract}

\maketitle

\section{Introduction}
\label{sec:intro}

Seed mutations in a cluster algebra produce new seeds from old ones, each of which is a tuple of the exchange matrix, the coefficients and the cluster variables, via their birational equations called the exchange relations \cite{FZ02}.
A cluster algebra of rank $r$ has the seed mutations in $r$ directions, hence the network of seeds generated by iteration of the seed mutations from the initial one forms an $r$-regular tree in which every vertex (seed) is connected with exactly $r$ vertices respectively by an edge (mutation).
The rank of a cluster algebra is defined to be the number of cluster variables in the initial seed.
In order to find significant paths (sequences of seeds) in such huge network consisting of infinitely many seeds, we often use the periodicity as an important indicator.
Fordy and Marsh defined the cluster mutation-periodic quivers concerning periodicity of the exchange matrices and discussed their dynamical properties under the seed mutations \cite{FM11}.
By using the notion of cluster mutation-periodic quivers, we relate seed mutations in cluster algebras with dynamical systems governed by birational maps and investigate the seed mutations in terms of the methods of dynamical systems. 
Unfortunately, almost all of the infinitely many dynamical systems thus related with cluster algebras do not have integrable structures; nevertheless, we can find abundant integrable systems among them.
In fact, since the introduction of cluster algebras by Fomin and Zelevinsky in 2002 \cite{FZ02} we have found plenty of cluster algebras related with discrete/quantum integrable systems such as discrete soliton equations, integrable maps on algebraic curves, discrete/$q$- Painlev\'e equations and Y-systems \cite{FZ03-2,IIKNS10,IIKKN13,IIKKN13-2,Keller13,Okubo13,Mase13,Okubo15,Marshakov13,Mase16,Nobe16,BGM18}.
Thus we see that appropriate paths in the network of seeds in adequate cluster algebras are strongly related with integrable systems, and hence it is expected that we find unknown integrable systems among cluster algebras.

It is well known that appropriate paths in the network of seeds in a cluster algebra can be assigned to the generalized Cartan matrices (GCMs)  via the exchange matrices in the paths \cite{FZ02,FZ03,FZ07}.
Especially, a rank 2 cluster algebra itself is assigned to a $2\times2$ GCM since it has the unique non-trivial path in the network of seeds, and the cluster algebra is referred to as of the GCM type.
In the preceding papers \cite{Nobe16,NM18,Nobe19}, the authors investigated a certain family of rank 2 cluster algebras from the viewpoint of discrete integrability.
The family consists of infinitely many cluster algebras some of which have integrable structures and are respectively assigned to the GCMs of finite and affine types.
The remaining infinitely many members in the family, however, are non-integrable and are assigned to the GCMs of strictly hyperbolic type.
We saw the integrability via conserved quantities of the dynamical systems governed by birational maps derived from the seed mutations.
Moreover, the integrable systems associated with the rank 2 cluster algebras of affine types, $A^{(1)}_1$ and $A^{(2)}_2$, have linear degree growth of the map iteration \cite{BV99}, and hence the systems are  linearizable \cite{DF01,RGSM11}.
Based on these results, in this paper, we consider the seed mutations of rank $N+1$ assigned to the GCM of type $A^{(1)}_N$ for $N\geq2$.
It should be noted that, in general, a sequence of seed mutations in a cluster algebra of higher rank is not assigned to any GCM in contrast to the case of rank 2 mentioned above because the GCMs of the exchange matrices in the sequence of higher rank seed mutations are not unique.
Nevertheless, the sequence of seed mutations considered in this paper is so carefully chosen from the network of seeds in a certain cluster algebra of rank $N+1$ that it can be assigned to the unique GCM of type $A^{(1)}_N$.
We note that the sequence of seed mutations thus chosen has several remarkable periodicities; the quivers associated with the exchange matrices are the cluster mutation-periodic quivers with period 1 \cite{FM11}, the coefficients generate certain $N$ monomials periodic with period $N$ and the cluster variables also generate certain $N$ Laurent polynomials periodic with the same period $N$ under iteration of the seed mutations.
By using the periodicity of exchange matrices, we respectively obtain dynamical systems of the coefficients and of the cluster variables from the sequence of seed mutations assigned to the GCM of type $A^{(1)}_N$.
Sufficiently many conserved quantities of each dynamical system naturally follow from the monomials and the Laurent polynomials, both of which have the same periodicity, respectively.
Moreover, the dynamical system of cluster variables is non-autonomously linearized by virtue of the periodic Laurent polynomials similar to the rank 2 cases investigated in the previous papers \cite{Nobe16, Nobe19}.
Due to the linearizability, the general solution to the dynamical system is concretely constructed, and it gives the general terms of the cluster variables.

This paper is organized as follows.
In \S \ref{sec:CABM}, we briefly review cluster algebras.
We then introduce the sub-cluster pattern which assigns the sequence of seed mutations to the path referred to as of type $A^{(1)}_N$ in the $(N+1)$-regular tree.
In \S \ref{sec:dynamicsofcoefficients}, we deduce periodicity of certain $N$ monomials generated by the coefficients assigned to the path of type $A^{(1)}_N$.
By using the monomials with periodcity, we obtain the general terms of the coefficients.
Then, in \S \ref{sec:dynamicsofclustervariables}, we introduce the dynamical system of cluster variables assigned to the path of type $A^{(1)}_N$. 
We also deduce periodicity of certain $N$ Laurent polynomials generated by the cluster variables.
Since the $N$ Laurent polynomials generate the $N$ functionally independent conserved quantities of the dynamical system of $N+1$ variables, it suggests the system to be integrable.
Moreover, the Laurent polynomials non-autonomously linearize the dynamical system and its general solution follows immediately.
\S \ref{sec:CONCL} is devoted to concluding remarks.
In Appendix \ref{sec:Appbirationalmap}, we consider the dynamical system investigated in \S \ref{sec:dynamicsofclustervariables} in the projective space $\P^{N+1}(\C)$, and induce the invariant curve of the system.

\section{Seed mutations of type $\boldsymbol{A^{(1)}_N}$}
\label{sec:CABM}

Let us introduce the seed $(\bx,\by,B)$, where we refer to $\bx=(x_1,x_2,\ldots,x_n)$ as the cluster of the seed, to $\by=(y_1,y_2,\ldots,y_n)$ as the coefficient tuple and to $B=(b_{ij})$ as the exchange matrix.
The number $n$ of variables in $\bx$ is called the rank of the seed.
The field $\mathcal{F}=\mathbb{QP}(\bx)$ generated by the cluster $\bx$ is referred to as the ambient field, where $\P=\left(\P,\cdot,\oplus\right)$ is a semifield endowed with multiplication $\cdot$ and auxiliary addition $\oplus$ and $\mathbb{QP}$ is the group ring of $\P$ over $\Q$.
The coefficient tuple $\by$ is taken from $\P^n$ and the exchange matrix $B$ is an $n\times n$ skew-symmetrizable integral matrix \cite{FZ02,FZ03,FZ07}.

Next we introduce the seed mutations.
For an integer $k\in\{1,2,\ldots,n\}$, the seed mutation $\mu_k$ transforms a seed $(\bx,\by,B)$ into the seed $(\bx^\prime,\by^\prime,B^\prime):=\mu_k(\bx,\by,B)$ defined by the following birational equations called the exchange relations:
\begin{align}
b_{ij}^\prime
&=
\begin{cases}
-b_{ij}&\mbox{$i=k$ or $j=k$},\\
b_{ij}+[-b_{ik}]_+b_{kj}+b_{ik}[b_{kj}]_+&\mbox{otherwise},\\
\end{cases}
\label{eq:mutem}\\
y_j^\prime
&=
\begin{cases}
(y_k)^{-1}&\mbox{$j=k$},\\
y_jy_k^{[b_{kj}]_+}(y_k\oplus 1)^{-b_{kj}}&\mbox{$j\neq k$},\\
\end{cases}
\label{eq:mutcoef}\\
x_j^\prime
&=
\begin{cases}
\DIS\frac{\DIS y_k\prod_{i=1}^n x_i^{[b_{ik}]_+}+\prod_{i=1}^n x_i^{[-b_{ik}]_+}}{(y_k\oplus 1)x_k}&\mbox{$j=k$},\\
x_j&\mbox{$j\neq k$},\\
\end{cases}
\label{eq:mutcv}
\end{align}
where we define $[a]_+:=\max[a,0]$ for $a\in\Z$.

Let $\T_n$ be the $n$-regular tree whose edges are labeled by the integers $1, 2, \ldots, n$ so that the $n$ edges emanating from each vertex receive different labels.
We write $t\ \overset{k}{\begin{xy}\ar @{-}(10,0)\end{xy}}\ t^\prime$ to indicate that vertices $t,t^\prime\in\T_n$ are joined by an edge labeled by $k$.
We assign a seed $\Sigma_t=(\bx_t,\by_t,B_t)$ to every vertex $t\in\T_n$ so that the seeds assigned to the endpoints of any edge $t\ \overset{k}{\begin{xy}\ar @{-}(10,0)\end{xy}}\ t^\prime$ are obtained from each other by the seed mutation $\mu_k$.
We refer to the assignment $\T_n\ni t\mapsto\Sigma_t$ as a cluster pattern.
We write the elements of the seed $\Sigma_t$ as follows
\begin{align*}
\bx_t=(x_{1;t},x_{2;t},\ldots,x_{n;t}),\quad
\by_t=(y_{1;t},y_{2;t},\ldots,y_{n;t}),\quad
B_t=(b_{ij}^t).
\end{align*}

Given a cluster pattern $\T_n\ni t\mapsto\Sigma_t$, we denote the union of clusters of all seeds in the pattern by
\begin{align*}
\mathcal{X}
:=
\bigcup_{t\in\T_n}\bx_t
=
\left\{
x_{i;t}\ |\ t\in\T_n,\ 1\leq i\leq n
\right\}.
\end{align*}
The cluster algebra $\mathcal{A}=\mathbb{ZP}[\mathcal{X}]$ associated with the cluster pattern is the $\mathbb{ZP}$-subalgebra of the ambient field $\mathcal{F}$ generated by all cluster variables.
It is well known that $\mathcal{A}$ is also generated by its initial cluster variables $x_1,x_2,\ldots,x_n$ as the Laurent polynomial subring of the ambient field $\mathcal{F}$ \cite{FZ02}.

Now we introduce the seed mutations assigned to the GCM of type $A^{(1)}_N$.
Let us consider the following initial seed $\Sigma_0=\left(\bx_0,\by_0,B_0\right)$ of rank $N+1$:
\begin{align*}
\bx_0
&=
\left(x_1,x_2,\cdots,x_{N+1}\right),
\\
\by_0
&=
\left(y_1,y_2,\cdots,y_{N+1}\right),
\\
B_0
&=
\left(
\begin{matrix}
0&-1&0&\cdots&0&-1\\
1&0&-1&\ddots&&0\\
0&1&\ddots&\ddots&\ddots&\vdots\\
\vdots&\ddots&\ddots&\ddots&-1&0\\
0&&\ddots&1&0&-1\\
1&0&\dots&0&1&0\\
\end{matrix}
\right),
\end{align*}
where the $(i,j)$-element $b_{ij}$ of the skew-symmetrix matrix $B_0$ is defined to be
\begin{align*}
b_{ij}
=
\begin{cases}
1&\mbox{$(i,j)=(N+1,1)$ or $(i,j)=(i,i-1)$ for $2\leq i\leq N+1$},\\
-1&\mbox{$(i,j)=(1,N+1)$ or $(i,j)=(i,i+1)$ for $1\leq i\leq N$},\\
0&\mbox{otherwise}.
\end{cases}
\end{align*}

We assume $\P$ to be the tropical semifield $\left({\rm Trop}(\by_0),\cdot,\oplus\right)$ generated by $\by_0$ \cite{FZ07}.
The multiplication $\cdot$ and the auxiliary addition $\oplus$ in $\P$ are respectively defined as follows
\begin{align*}
y_1^{a_1}y_2^{a_2}\cdots y_n^{a_n}
\cdot
y_1^{b_1}y_2^{b_2}\cdots y_n^{b_n}
&:=
y_1^{a_1+b_1}y_2^{a_2+b_2}\cdots y_n^{a_n+b_n},
\\
y_1^{a_1}y_2^{a_2}\cdots y_n^{a_n}
\oplus
y_1^{b_1}y_2^{b_2}\cdots y_n^{b_n}
&:=
y_1^{\min[a_1,b_1]}y_2^{\min[a_2,b_2]}\cdots y_n^{\min[a_n,b_n]}
\end{align*}
for $a_i,b_i\in\Z$ ($i=1,2,\ldots,n$).

We give a cluster pattern.
Let  $\T_{N+1}$  be the $(N+1)$-regular tree whose edges are labeled by $1,2,\ldots,N+1$. 
We label the vertices in $\T_{N+1}$ in the following manner.
First choose an arbitrary vertex and denote it by $t_0$, which is assigned to the initial seed $\Sigma_0$.
Next denote the vertex connected with $t_0$ by the edge labelled by $1$ by $t_1$.
The vertex $t_1$ is assigned to the seed $\Sigma_1$ obtained from $\Sigma_0$ by applying the seed mutation $\mu_1$.
Then inductively denote the vertex connected with $t_{\ell(N+1)+k-1}$ by the edge labeled by $k$ by $t_{\ell(N+1)+k}$ for $\ell\geq0$ and $k=1,2,\ldots, N+1$.
Since the vertex $t_{\ell(N+1)+k}$ is assigned to the seed $\Sigma_{\ell(N+1)+k}$, the seed $\Sigma_{\ell(N+1)+k}$ is obtained from $\Sigma_0$ by applying the following sequence of seed mutations
\begin{align*}
\underbrace{
\underbrace{\mu_1,\mu_2,\ldots,\mu_{N+1}}_{N+1},
\underbrace{\mu_1,\mu_2,\ldots,\mu_{N+1}}_{N+1},
\ldots,
\underbrace{\mu_1,\mu_2,\ldots,\mu_{N+1}}_{N+1}}_{\ell\times(N+1)},
\mu_1,\mu_2,\ldots,\mu_k.
\end{align*}
Thus we obtain the path $(t_0,t_1,t_2,\ldots)$ in the tree $\T_{N+1}$ (see figure \ref{fig:binarytree}) and denote it by $\varpi$.
\begin{figure}[htbp]
\begin{align*}
\xymatrix@C=20pt{
t_0
\ar@{-}[r]^-{1}
&
t_1
\ar@{-}[r]^-{2}
&
\cdots
&
t_{N+1}
\ar@{-}[l]_-{N+1}
\ar@{-}[r]^-{1}
&
t_{(N+1)+1}
\ar@{-}[r]^-{2}
&
\cdots\cdots\cdots
&
t_{\ell(N+1)-1}
\ar@{-}[l]_-{N}
\\
&
&
\cdots
\ar@{-}[r]^-{1}
&
t_{(\ell+1)(N+1)}
\ar@{-}[r]^-{N+1}
&
\cdots
&
t_{\ell(N+1)+1}
\ar@{-}[l]_-{2}
&
t_{\ell(N+1)}
\ar@{-}[u]_-{N+1}
\ar@{-}[l]_-{1}
}
\end{align*}
\caption{
The path $\varpi$ in the $(N+1)$-regular tree $\T_{N+1}$.
}
\label{fig:binarytree}
\end{figure}

Let the set of seeds be
\begin{align*}
\bSigma:=\left\{\Sigma_0,\Sigma_1,\Sigma_2,\ldots\right\}.
\end{align*}
Then we obtain the partial assignment $\T_{N+1}\supset\varpi\to\bSigma$;
\begin{align*}
t_{\ell(N+1)+k}\mapsto\Sigma_{\ell(N+1)+k}
\end{align*}
for $\ell\geq0$ and $k=1,2,\ldots, N+1$.
We call the partial assignment the sub-cluster pattern, and fix it throughout this paper.
Note that we need not whole cluster pattern but the sub-cluster pattern for our purpose.

The quiver $Q_0$ associated with the exchange matrix $B_0$ assigned to the vertex $t_0$ in the path $\varpi$ is given as follows
\begin{align*}
\xymatrix{
Q_0
&
&&
\overset{1}{\bigotimes}
& 
\\
&
\overset{2}{\bigcirc}
\ar@{->}[rru]
& 
\overset{3}{\bigcirc}
\ar@{->}[l]
&
\cdots
\ar@{->}[l]
&
\overset{N}{\bigcirc}
\ar@{->}[l]
&
\overset{N+1}{\bigodot}.
\ar@{->}[llu]
\ar@{->}[l]
}
\end{align*}
The vertex labeled by $1$ is a sink, denoted by $\bigotimes$, and the one by $N+1$ is a source, denoted by $\bigodot$.
The quiver $Q_1=\mu_1(Q_0)$ associated with the exchange matrix $B_1=\mu_1(B_0)$ assigned to the vertex $t_1$ in $\varpi$ is obtained by reversing the arrows connected with the vertex 1 in $Q_0$ (see \eqref{eq:mutem}).
Note that, in $Q_1$, the vertex 2 is a sink and the vertex 1 is a source:
\begin{align*}
\xymatrix{
Q_1
&
&&
\overset{2}{\bigotimes}
& 
\\
&
\overset{3}{\bigcirc}
\ar@{->}[rru]
& 
\overset{4}{\bigcirc}
\ar@{->}[l]
&
\cdots
\ar@{->}[l]
&
\overset{N+1}{\bigcirc}
\ar@{->}[l]
&
\overset{1}{\bigodot}.
\ar@{->}[llu]
\ar@{->}[l]
}
\end{align*}
It is easy to see that we inductively obtain the quiver $Q_{\ell(N+1)+k}$ associated with the exchange matrix $B_{\ell(N+1)+k}$ assigned to the vertex $t_{\ell(N+1)+k}$ in $\varpi$ as follows
\begin{align*}
\xymatrix{
Q_{\ell(N+1)+k}
&
&&
\overset{k+1}{\bigotimes}
& 
\\
&
\overset{k+2}{\bigcirc}
\ar@{->}[rru]
& 
\overset{k+3}{\bigcirc}
\ar@{->}[l]
&
\cdots
\ar@{->}[l]
&
\overset{k-1}{\bigcirc}
\ar@{->}[l]
&
\overset{k}{\bigodot},
\ar@{->}[llu]
\ar@{->}[l]
}
\end{align*}
where the labels are reduced modulo $N+1$.

Thus, for $\ell\geq0$ and $k=1,2,\ldots, N+1$, we obtain the periodicity of the quivers 
\begin{align*}
Q_{\ell(N+1)+k}
=
\left(\sigma_{N+1}\right)^kQ_0,
\end{align*}
where $\sigma_{N+1}\in\mathfrak{S}_{N+1}$ is the permutation
\begin{align}
\sigma_{N+1}
=
\left(
\begin{matrix}
1&2&\cdots&N&N+1\\
N+1&1&\cdots&N-1&N\\
\end{matrix}
\right)
\label{eq:sigmaN+1}
\end{align}
of $N+1$ letters.
Remark that the action $\tau Q$ of the permutation $\tau\in\mathfrak{S}_{N+1}$ on the quiver $Q$ with $N+1$ vertices is, in general, defined as follows \cite{FM11}
\begin{align*}
&\sharp
\left\{
\mbox{arrows emanating from the vertex $i\in Q$ to $j\in Q$}
\right\}
\\
&=
\sharp
\left\{
\mbox{arrows emanating from the vertex $(\tau)^{-1}(i)\in \tau Q$ to $(\tau)^{-1}(j)\in \tau Q$}
\right\}.
\end{align*}
Thus the quivers $Q_0,Q_1,Q_2,\ldots$ are the cluster mutation-periodic quivers with period 1 \cite{FM11}.

The exchange matrices also have the same periodicity as the quivers:
\begin{align*}
B_{\ell(N+1)+k}
=
\left(\sigma_{N+1}\right)^kB_0.
\end{align*}
The action $\sigma_{N+1}B_m$ of $\sigma_{N+1}$ on the exchange matrix $B_m$ is defined by using the permutation matrix \cite{FM11} so that it is compatible with the correspondence between $B_m$ and $Q_m$.

For any $m\geq0$, the Cartan counterpart \cite{FZ03} $A(B_m)$ of the exchange matrix $B_m$ is given as
\begin{align*}
A(B_m)
:=
\left(2\delta_{ij}-\left|b_{ij}^m\right|\right)
=
\left(
\begin{matrix}
2&-1&0&\cdots&0&-1\\
-1&2&-1&\ddots&&0\\
0&-1&\ddots&\ddots&\ddots&\vdots\\
\vdots&\ddots&\ddots&\ddots&-1&0\\
0&&\ddots&-1&2&-1\\
-1&0&\dots&0&-1&2\\
\end{matrix}
\right).
\end{align*}
Since $A(B_m)$ is the GCM of type $A^{(1)}_N$, we refer to the path $\varpi$ in the tree $\T_{N+1}$ assigned to the set $\bSigma$ of seeds as of type $A^{(1)}_N$.

A (skew-) symmetrizable matrix $A$ is a matrix that can be written as $A=DS$, where $D$ is a diagonal matrix and $S$ is a (skew-) symmetric matrix.
A GCM is a symmetrizable matrix $A=(a_{ij})$ with integral entries such that (1) $a_{ii}=2$, (2) $a_{ij}\leq0$ for $i\neq j$ and (3) $a_{ij}=0$ if and only if $a_{ji}=0$.
An indecomposable GCM $A$ is said to be of finite type if its principal minors are positive, of affine type if its proper principal minors are positive and $\det A=0$, and of indefinite type otherwise \cite{Kac}.
A cluster algebra $\mathcal{A}=\mathbb{ZP}[\mathcal{X}]$ is said to be of finite type if the set of seeds $\mathcal{X}$ is finite.
It is well known that there is a canonical bijection between the GCMs of finite type and the strong isomorphism classes of series of cluster algebras of finite type. 
Under this bijection, a GCM $A$ of finite type corresponds to the series $A(B,-)$, where $B$ is an arbitrary skew-symmetric matrix with $A(B)=A$ \cite{FZ03}.

\section{Dynamics of coefficients}
\label{sec:dynamicsofcoefficients}
We consider dynamics of the coefficients assigned to the path $\varpi$ of type $A^{(1)}_N$.
In order to analyze the dynamics of coefficients, we first consider periodicity of certain monomials generated by the coefficients.

\subsection{Periodicity}
\label{subsec:pcoef}
First we show a lemma concerning the coefficient tuple $\by_{N+1}=\left(\mu_{N+1}\circ\cdots\circ\mu_2\circ\mu_1\right)(\by_0)$ obtained by applying the consecutive seed mutations $\mu_{N+1}\circ\cdots\circ\mu_2\circ\mu_1$ to the initial one $\by_0=(y_1,y_2,\ldots,y_N)$.

\begin{lemma}\label{lem:y1}
Let $\by_0=(y_1,y_2,\ldots,y_{N+1})$ be the initial coefficient tuple.
Then, for any $N\geq2$, we have
\begin{align*}
\by_{N+1}
=
(y_{1;N+1},y_{2;N+1},\ldots,y_{N+1;N+1})
=
\left(
(y_1)^{-1},(y_2)^{-1},\ldots,(y_{N+1})^{-1}
\right).
\end{align*}
\end{lemma}

(Proof)\quad
By using the exchange 
relation
\eqref{eq:mutcoef} for the coefficients, we inductively compute
\begin{align*}
\by_0
&=
\left(
y_1,y_2,\ldots,y_{N+1}
\right)
\\
\overset{\mu_1}{\longleftrightarrow}
\by_1
&=
\left(
(y_1)^{-1},y_2(y_1\oplus1),y_3,\ldots,y_N,y_{N+1}(y_1\oplus1)
\right)
\\
&=
\left(
(y_1)^{-1},y_2,y_3,\ldots,y_N,y_{N+1}
\right)
\\
\overset{\mu_2}{\longleftrightarrow}
\by_2
&=
\left(
(y_1)^{-1}(y_2\oplus1),(y_2)^{-1},y_3(y_1\oplus1),y_4,\ldots,y_{N+1}
\right)
\\
&=
\left(
(y_1)^{-1},(y_2)^{-1},y_3,y_4,\ldots,y_{N+1}
\right)
\\
&\cdots
\\
\overset{\mu_{N+1}}{\longleftrightarrow}
\by_{N+1}
&=
\left(
(y_1)^{-1}(y_{N+1}\oplus1),(y_2)^{-1},\ldots,(y_{N-1})^{-1},(y_N)^{-1}(y_{N+1}\oplus1),(y_{N+1})^{-1}
\right)
\\
&=
\left(
(y_1)^{-1},(y_2)^{-1},\ldots,(y_{N-1})^{-1},(y_N)^{-1},(y_{N+1})^{-1}
\right),
\end{align*}
where we use the fact $y_j\oplus1=y_j^{\min[1,0]}=1$ for $j=1,2,\ldots,N+1$.
\qed

Now we consider iteration of the consecutive seed mutations $\mu_{N+1}\circ\cdots\circ\mu_2\circ\mu_1$, which defines the map $\by_{\ell(N+1)}\mapsto \by_{(\ell+1)(N+1)}$ for $\ell\geq1$.
Let us introduce the monomials $\nu_1^\ell,\nu_2^\ell,\ldots,\nu_N^\ell$ in the tropical semifield $\P=\left({\rm Trop}(\by_0),\cdot,\oplus\right)$
\begin{align*}
\nu_1^\ell
&:=
y_{N+1;\ell(N+1)}y_{1;\ell(N+1)},
\\
\nu_j^\ell
&:=
y_{j;\ell(N+1)}
\quad
(j=2,3,\ldots,N)
\end{align*}
generated by the coefficients in
\begin{align*}
\by_{\ell(N+1)}
=
\left(
y_{1;\ell(N+1)},y_{2;\ell(N+1)},\ldots,y_{N+1;\ell(N+1)}
\right).
\end{align*}

The permutation
\begin{align}
\sigma_N
=
\left(
\begin{matrix}
1&2&\cdots&N-1&N\\
N&1&\cdots&N-2&N-1\\
\end{matrix}
\right)\in\mathfrak{S}_N
\label{eq:sigmaN}
\end{align}
 of $N$ letters acts on the monomial $\nu_j^\ell$ as
\begin{align*}
\sigma_N\nu_j^\ell
=
\nu_{(\sigma_N)^{-1}(j)}^\ell
=
\nu_{j+1}^\ell
\end{align*}
for $j=1,2,\ldots,N$, where the subscript is reduced modulo $N$.

\begin{thm}\label{thm:coefgcq}
For any $\ell\geq1$ and $j=1,2,\ldots,N$, we have
\begin{align*}
\nu_j^{\ell+1}
=
\sigma_N\nu_j^\ell
=
\nu_{j+1}^\ell,
\end{align*}
where the subscript is reduced modulo $N$.
Therefore, every $\nu_j^\ell$ has period $N$ on $\ell$:
\begin{align*}
\nu_j^{\ell+N}
=
\nu_j^\ell.
\end{align*}
\end{thm}

(Proof)\quad
First note that the exchange matrix $B_k$ has the periodicity
\begin{align*}
B_{\ell(N+1)+k}=B_k
\end{align*}
for $\ell\geq0$ and $k=1,2,\ldots,N+1$.
We then see that the $k$-th row of the exchange matrix $B_{\ell(N+1)+k-1}=B_{k-1}$ for $k=1,2,\ldots,N+1$, which determines the seed mutation $\mu_k$, has two non-zero elements $-1$ at $(k,k-1)$ and at $(k,k+1)$:
\begin{align*}
\bordermatrix{
&1&\cdots&k-2&k-1&k&k+1&k+2&\cdots&N+1\cr
k&0&\cdots&0&-1&0&-1&0&\cdots&0\cr
}.
\end{align*}
Thus the exchange relation \eqref{eq:mutcoef} reduces to
\begin{align}
y_{j;(N+1)+k}
&=
\mu_k(y_{j;(N+1)+k-1})
=
\begin{cases}
(y_{k;(N+1)+k-1})^{-1}&\mbox{$j=k$},\\
y_{j;(N+1)+k-1}(y_{k;(N+1)+k-1}\oplus 1)&\mbox{$j=k\pm1$},\\
y_{j;(N+1)+k-1}&\mbox{$j\neq k,k\pm1$}.\\
\end{cases}
\label{eq:ercoefnew}
\end{align}
Also note that, by lemma \ref{lem:y1}, we have
\begin{align*}
\by_{N+1}
=
\left(
y_{1;N+1},y_{2;N+1},\ldots,y_{N+1;N+1}
\right)
=
\left(
(y_1)^{-1},(y_2)^{-1},\ldots,(y_{N+1})^{-1}
\right).
\end{align*}

Apply the mutation $\mu_1$ to $\by_{N+1}$.
We then obtain
\begin{align*}
y_{1;(N+1)+1}
&=
\mu_1(y_{1;N+1})
=
(y_{1;N+1})^{-1},
\nonumber\\
y_{2;(N+1)+1}
&=
\mu_1(y_{2;N+1})
=
y_{2;N+1}(y_{1;N+1}\oplus1)
=
y_{2;N+1}y_{1;N+1},
\\
y_{j;(N+1)+1}
&=
\mu_1(y_{j;N+1})
=
y_{j;N+1}
\quad
(j=3,4,\ldots,N),
\nonumber\\
y_{N+1;(N+1)+1}
&=
\mu_1(y_{N+1;N+1})
=
y_{N+1;N+1}(y_{1;N+1}\oplus1)
=
y_{N+1;N+1}y_{1;N+1},
\nonumber
\end{align*}
where we use the exchange relation \eqref{eq:ercoefnew} and the fact $y_{1;N+1}=(y_1)^{-1}$ which implies $y_{1;N+1}\oplus1=y_{1;N+1}$.

We inductively obtain
\begin{align*}
y_{j;(N+1)+k}
&=
\mu_k(y_{j;(N+1)+k-1})
=
y_{j;(N+1)+k-1}
=
y_{j+1;N+1}
\quad
(j=1,2,\ldots,k-2),\\
y_{k-1;(N+1)+k}
&=
\mu_k(y_{k-1;(N+1)+k-1})
=
y_{k-1;(N+1)+k-1}y_{k;(N+1)+k-1}
=
y_{k;N+1},
\\
y_{k;(N+1)+k}
&=
\mu_k(y_{k;(N+1)+k-1})
=
(y_{k;(N+1)+k-1})^{-1}
=
(y_{1;N+1}y_{2;N+1}\cdots y_{k;N+1})^{-1},
\\
y_{k+1;(N+1)+k}
&=
\mu_k(y_{k+1;(N+1)+k-1})
=
y_{k+1;(N+1)+k-1}y_{k;(N+1)+k-1}
=
y_{1;N+1}y_{2;N+1}\cdots y_{k+1;N+1},
\\
y_{j;(N+1)+k}
&=
\mu_k(y_{j;(N+1)+k-1})
=
y_{j;(N+1)+k-1}
=
y_{j;N+1}
\quad
(j=k+2,k+3,\ldots,N),\\
y_{N+1;(N+1)+k}
&=
\mu_k(y_{N+1;(N+1)+k-1})
=
y_{N+1;(N+1)+k-1}
=
y_{1;N+1}y_{N+1;N+1}
\end{align*}
by applying the consecutive seed mutations $\mu_k\circ\cdots\circ\mu_3\circ\mu_2$ to $\by_{(N+1)+1}$ ($k=2,3,\ldots,N-1$).
Note that we use the fact
\begin{align}
y_{k;(N+1)+k-1}
=
y_{1;N+1}y_{2;N+1}\cdots y_{k;N+1}
=
(y_1)^{-1}(y_2)^{-1}\cdots (y_k)^{-1}
\label{eq:tropsum}
\end{align}
for $k=2,3,\ldots,N-1$ which implies $y_{k;(N+1)+k-1}\oplus1=y_{k;(N+1)+k-1}$.

Moreover, apply $\mu_N$ to $\by_{(N+1)+N-1}$. 
Then we have
\begin{align}
y_{j;(N+1)+N}
&=
\mu_N(y_{j;(N+1)+N-1})
=
y_{j;(N+1)+N-1}
=
y_{j+1;N+1}
\quad
(j=1,2,\ldots,N-2),
\nonumber\\
y_{N-1;(N+1)+N}
&=
\mu_N(y_{N-1;(N+1)+N-1})
=
y_{N-1;(N+1)+N-1}y_{N;(N+1)+N-1}
=
y_{N;N+1},
\nonumber\\
y_{N;(N+1)+N}
&=
\mu_N(y_{N;(N+1)+N-1})
=
(y_{N;(N+1)+N-1})^{-1}
=
(y_{1;N+1}y_{2;N+1}\cdots y_{N;N+1})^{-1},
\nonumber\\
y_{N+1;(N+1)+N}
&=
\mu_N(y_{N+1;(N+1)+N-1})
=
y_{N+1;(N+1)+N-1}y_{N;(N+1)+N-1}
\nonumber\\
&=
(y_{1;N+1})^2y_{2;N+1}\cdots y_{N+1;N+1},
\label{eq:tropsumN+1}
\end{align}
where we use \eqref{eq:tropsum} for $k=N$ which implies $y_{N;(N+1)+N-1}\oplus1=y_{N;(N+1)+N-1}$.
Finally, by applying $\mu_{N+1}$ to $\by_{(N+1)+N}$, we obtain
\begin{align*}
y_{1;2(N+1)}
&=
\mu_{N+1}(y_{1;(N+1)+N})
=
y_{1;(N+1)+N}y_{N+1;(N+1)+N}
\\
&=
(y_{1;N+1})^2(y_{2;N+1})^2y_{3;N+1}\cdots y_{N+1;N+1},
\\
y_{j;2(N+1)}
&=
\mu_{N+1}(y_{j;(N+1)+N})
=
y_{j;(N+1)+N}
=
y_{j+1;N+1}
\quad
(j=2,3,\ldots,N-1),
\\
y_{N;2(N+1)}
&=
\mu_{N+1}(y_{N;(N+1)+N})
=
y_{N;(N+1)+N}y_{N+1;(N+1)+N}
=
y_{1;N+1}y_{N+1;N+1},
\\
y_{N+1;2(N+1)}
&=
\mu_{N+1}(y_{N+1;(N+1)+N})
=
(y_{N+1;(N+1)+N})^{-1}
=
\left((y_{1;N+1})^2y_{2;N+1}\cdots y_{N+1;N+1}\right)^{-1},
\end{align*}
where we use \eqref{eq:tropsumN+1} which implies $y_{N+1;(N+1)+N}\oplus1=y_{N+1;(N+1)+N}$.
It immediately follows
\begin{align*}
\nu_1^2
&=
y_{N+1;2(N+1)}y_{1;2(N+1)}
=
y_{2;N+1}
=
\nu_2^1,
\\
\nu_j^2
&=
y_{j;2(N+1)}
=
y_{j+1;N+1}
=
\nu_{j+1}^1
\quad
(j=2,3,\ldots,N-1),
\\
\nu_N^2
&=
y_{N;2(N+1)}
=
y_{1;N+1}y_{N+1;N+1}
=
\nu_1^1.
\end{align*}

Every coefficient $y_{j;2(N+1)}$ in $\by_{2(N+1)}$ except for $y_{N+1;2(N+1)}$ is a monomial consisting of negative powers of the initial ones $y_1,y_2,\ldots,y_{N+1}$.
Thus the exchange relation \eqref{eq:ercoefnew} reduces to
\begin{align}
y_{j;\ell(N+1)+k}
&=
\mu_k(y_{j;\ell(N+1)+k-1})
=
\begin{cases}
(y_{k;\ell(N+1)+k-1})^{-1}&\mbox{$j=k$},\\
y_{j;\ell(N+1)+k-1}y_{k;\ell(N+1)+k-1}&\mbox{$j=k\pm1$},\\
y_{j;\ell(N+1)+k-1}&\mbox{$j\neq k,k\pm1$}\\
\end{cases}
\label{eq:ercoeffnew}
\end{align}
for any $\ell\geq2$ as in the case with $\ell=1$.
Therefore, for any $\ell\geq2$, we inductively obtain
\begin{align*}
y_{1;(\ell+1)(N+1)}
&=
(y_{1;\ell(N+1)})^2(y_{2;\ell(N+1)})^2y_{3;\ell(N+1)}\cdots y_{N+1;\ell(N+1)},
\\
y_{j;(\ell+1)(N+1)}
&=
y_{j+1;\ell(N+1)}
\quad
(j=2,3,\ldots,N-1),
\\
y_{N;(\ell+1)(N+1)},
&=
y_{1;\ell(N+1)}y_{N+1;\ell(N+1)},
\\
y_{N+1;(\ell+1)(N+1)}
&=
\left((y_{1;\ell(N+1)})^2y_{2;\ell(N+1)}\cdots y_{N+1;\ell(N+1)}\right)^{-1},
\end{align*}
which implies
\begin{align*}
\nu_1^{\ell+1}
&=
y_{N+1;(\ell+1)(N+1)}y_{1;(\ell+1)(N+1)}
=
y_{2;\ell(N+1)}
=
\nu_2^\ell,
\\
\nu_j^{\ell+1}
&=
y_{j;(\ell+1)(N+1)}
=
y_{j+1;\ell(N+1)}
=
\nu_{j+1}^\ell
\quad
(j=2,3,\ldots,N-1),
\\
\nu_N^{\ell+1}
&=
y_{N;\ell(N+1)}
=
y_{1;\ell(N+1)}y_{N+1;\ell(N+1)}
=
\nu_1^\ell.
\end{align*}
It immediately follows $\nu_j^{\ell+N}=\nu_j^\ell$ for $\ell\geq1$ and $j=1,2,\ldots,N$.
\qed

We easily find the conserved quantities of the dynamics of coefficients via the monomials $\nu_1,\nu_2,\ldots,\nu_N$, where we denote $\nu_j^1$ by $\nu_j$ ($j=1,2,\ldots,N$) for simplicity.
Remark that we have
\begin{align}
\nu_1
&=
y_{N+1;N+1}y_{1;N+1}
=
(y_{N+1})^{-1}(y_1)^{-1},
\label{eq:nu1ef}\\
\nu_j
&=
y_{j;N+1}
=
(y_j)^{-1}
\quad
(j=2,,3,\ldots,N).
\label{eq:nujef}
\end{align}
In the tropical semifield $\P$, let $e_n$ ($n=1,2,\ldots,N$) be the fundamental symmetric polynomial of degree $n$ generated by the monomials $\nu_1,\nu_2,\ldots,\nu_N$:
\begin{align*}
e_n
=
e_n(\nu_1,\nu_2,\ldots,\nu_N)
:=
\bigoplus_{\substack{I\subset\left\{1,2,\ldots,N\right\}\\|I|=n}}\prod_{j\in I}\nu_j.
\end{align*}
We denote $e_n(\nu_1^\ell,\nu_2^\ell,\ldots,\nu_N^\ell)$ simply by $e_n^\ell$ for $\ell\geq2$.

For $\ell\geq1$, let us introduce new variables:
\begin{align*}
&\by^\ell
:=
\left(
y_1^\ell,y_2^\ell,\ldots,y_{N+1}^\ell
\right),
\\
&y_j^\ell:=y_{j;\ell (N+1)}
\quad
(j=1,2,\ldots,N+1).
\end{align*}
We see from the proof of theorem \ref{thm:coefgcq} that the evolution of $\by^\ell$ is given as follows
\begin{align}
\begin{cases}
y_1^{\ell+1}
=
\left(y_1^\ell y_2^\ell\right)^2y_3^\ell\cdots y_{N+1}^\ell,
\\
y_j^{\ell+1}
=
y_{j+1}^\ell
&
(j=2,3,\ldots,N-1),
\\
y_N^{\ell+1}
=
y_1^\ell y_{N+1}^\ell,
\\
y_{N+1}^{\ell+1}
=
\left(\left(y_1^\ell\right)^2y_2^\ell\cdots y_{N+1}^\ell\right)^{-1}.
\end{cases}
\label{eq:tecoef}
\end{align}
Thus the evolution of $\by^\ell$ defines the map $\psi:\P^{N+1}\to\P^{N+1};\by^\ell\mapsto\by^{\ell+1}$.

Hone and Inoue gave the recurrence formula (4.8) of the Y-system associated with a cluster mutation-periodic quiver with period 1 corresponding to a skew-symmetric matrix \cite{HI14}.
We remark that the evolution \eqref{eq:tecoef} of $\by^\ell$ reduces to a special case of their recurrence formula (4.8) by replacing auxiliary addition $\oplus$ with standard one $+$.

\begin{corollary}\label{cor:cqcoef}
The fundamental symmetric polynomial $e_n^\ell$ ($n=1,2,\ldots,N$) of degree $n$ generated by the monomials $\nu_1^\ell,\nu_2^\ell,\ldots,\nu_N^\ell$ is the conserved quantity of the dynamical system $\by^{\ell+1}=\psi(\by^\ell)$ governed by the map $\psi$, that is, we have
\begin{align*}
e_n^\ell=e_n
\end{align*}
for any $\ell\geq1$.
Moreover, all the fundamental symmetric polynomials are the same:
\begin{align*}
e_1=e_2=\cdots=e_N=\prod_{j=1}^{N+1}(y_j)^{-1}.
\end{align*}
\end{corollary}

(Proof)\quad
The permutation $\sigma_N$ acts on $e_n^\ell$ as follows
\begin{align*}
\sigma_Ne_n^\ell
&=
\sigma_Ne_n(\nu_1^\ell,\nu_2^\ell,\ldots,\nu_N^\ell)
\\
&=
e_n(\nu_{(\sigma_N)^{-1}(1)}^\ell,\nu_{(\sigma_N)^{-1}(2)}^\ell,\ldots,\nu_{(\sigma_N)^{-1}(N)}^\ell)
\\
&=
e_n(\nu_2^\ell,\nu_3^\ell,\ldots,,\nu_N^\ell,\nu_1^\ell)
\\
&=
e_n(\nu_1^{\ell+1},\nu_2^{\ell+1},\ldots,\nu_{N-1}^{\ell+1},\nu_N^{\ell+1})
=
e_n^{\ell+1},
\end{align*}
where we use $\nu_j^{\ell+1}=\nu_{j+1}^\ell$ for $j=1,2,\ldots,N$ (see theorem \ref{thm:coefgcq}).
On the other hand, the identity $\sigma_Ne_n^\ell=e_n^\ell$ holds by definition.

Moreover, by using \eqref{eq:nu1ef} and \eqref{eq:nujef}, we compute
\begin{align*}
e_1
&=
\nu_1\oplus\nu_2\oplus\cdots\oplus\nu_N
\\
&=
(y_1)^{-1}(y_{N+1})^{-1}\oplus (y_2)^{-1}\oplus\cdots\oplus (y_N)^{-1}
\\
&=
y_1^{\min[-1,0,\ldots,0]}y_2^{\min[0,-1,0,\ldots,0]}\cdots y_{N+1}^{\min[-1,0,\ldots,0]}
=
\prod_{j=1}^{N+1}(y_j)^{-1}.
\end{align*}
Similarly, since we have
\begin{align*}
\bigoplus_{\substack{I\subset\left\{2,3,\ldots,N\right\}\\|I|=n}}\prod_{j\in I}\nu_j
=
\bigoplus_{\substack{I\subset\left\{2,3,\ldots,N\right\}\\|I|=n}}\prod_{j\in I}(y_j)^{-1}
=
\prod_{j=2}^{N}(y_j)^{-1}
\end{align*}
for any $n\geq1$, we obtain
\begin{align*}
e_n
&=
\nu_1\bigoplus_{\substack{I\subset\left\{2,3,\ldots,N\right\}\\|I|=n-1}}\prod_{j\in I}\nu_j
\oplus
\bigoplus_{\substack{I\subset\left\{2,3,\ldots,N\right\}\\|I|=n}}\prod_{j\in I}\nu_j
\\
&=
(y_1)^{-1}(y_{N+1})^{-1}\prod_{j=2}^{N}(y_j)^{-1}
\bigoplus
\prod_{j=2}^{N}(y_j)^{-1}
=
\prod_{j=1}^{N+1}(y_j)^{-1}.
\end{align*}
This completes the proof.
\qed

\subsection{General solution}
\label{subsec:gscoef}
For $\ell=nN+s\geq1$ ($0\leq n$, $1\leq s\leq N$), we define the monomial $C_\ell$ in $\P$ to be
\begin{align*}
C_\ell
:=&
\nu_1\nu_2\cdots \nu_\ell
=
(e_1)^n(y_{N+1})^{-1}\prod_{j=1}^{s}(y_j)^{-1},
\end{align*}
where the subscript of $\nu_j$ is reduced modulo $N$ and we use \eqref{eq:nu1ef} and \eqref{eq:nujef}.
Then
\begin{align*}
C_N=e_1
\end{align*}
is the conserved quantity of the dynamical system $\by^{\ell+1}=\psi(\by^\ell)$. 

The dynamical system governed by the map $\psi:\by^\ell\mapsto\by^{\ell+1}$ is easily solved by using the monomials $C_1,C_2,\ldots,C_N$.

\begin{thm}\label{thm:cqgs}
For given $N\geq2$, put $\ell=n N+s$, where $0\leq n$ and $1\leq s\leq N$.
The general solution $\by^\ell=(y_1^\ell,y_2^\ell,\ldots,y_{N+1}^\ell)$ to the dynamical system $\by^{\ell+1}=\psi(\by^\ell)$ governed by the map $\psi$ is given by
\begin{align}
\begin{cases}
y_1^\ell
=
(C_N)^{n(N+1)+s-1}C_s y_{N+1},
\\[5pt]
\DIS
y_j^\ell
=
\nu_{j+s-1}
&
(j=2,3,\ldots,N),
\\[5pt]
y_{N+1}^\ell
=
\left((C_N)^{n(N+1)+s-1}C_{s-1}y_{N+1}\right)^{-1},
\\
\end{cases}
\label{eq:gsdscoef}
\end{align}
where the subscript of $\nu_j$ is reduced 
modulo 
$N$ and we assume $C_0=1$.
\end{thm}

(Proof)\quad
Since $y_j^\ell=\nu_j^\ell$ for $j=2,3,\ldots,N$, the solution 
\begin{align*}
y_j^\ell
&=
\nu_{j+s-1}
\quad
(j=2,3,\ldots,N)
\end{align*}
is a straightforward consequence of theorem \ref{thm:coefgcq}.

We compute $y_1^\ell$ and $y_{N+1}^\ell$.
For $\ell=1$, \textit{i.e.}, $n=0$ and $s=1$, we have
\begin{align*}
y_1^1
&=
(y_1)^{-1}
=
\nu_1y_{N+1}
=
(C_N)^0C_1y_{N+1},
\\
y_{N+1}^1
&=
(y_{N+1})^{-1}
=
\left(
(C_N)^0C_0y_{N+1}
\right)^{-1},
\end{align*}
where we use the assumption $C_0=1$ and \eqref{eq:nu1ef}.
We assume that \eqref{eq:gsdscoef} is true for $\ell=nN+s$.
Then, by using \eqref{eq:tecoef} and the fact that $C_N$ is the conserved quantity, we have
\begin{align*}
y_1^{\ell+1}
&=
\left(y_1^\ell y_2^\ell\right)^2y_3^\ell\cdots y_{N+1}^\ell
=
C_Ny_1^\ell y_2^\ell
\\
&=
C_N(C_N)^{n(N+1)+s-1}C_s y_{N+1}\nu_{s+1}
\\
&=
(C_N)^{n(N+1)+s}C_{s+1}y_{N+1},
\\
y_{N+1}^{\ell+1}
&=
\left(\left(y_1^\ell\right)^2y_2^\ell\cdots y_{N+1}^\ell\right)^{-1}
=
\left(C_Ny_1^\ell\right)^{-1}
\\
&=
\left(C_N(C_N)^{n(N+1)+s-1}C_s y_{N+1}\right)^{-1}
\\
&=
\left((C_N)^{n(N+1)+s}C_s y_{N+1}\right)^{-1}.
\end{align*}
Thus \eqref{eq:gsdscoef} is true for $\ell+1=nN+s+1$.
\qed

In order to compute the cluster mutation
\begin{align*}
\bx_{(\ell+1)(N+1)}=\left(\mu_{N+1}\circ\cdots\circ\mu_2\circ\mu_1\right)(\bx_{\ell(N+1)}),
\end{align*}
we use the exchange relation \eqref{eq:mutcv}.
Throughout the consecutive mutations $\mu_1,\mu_2,\ldots,\mu_{N+1}$, the cluster variable $x_{k;\ell(N+1)}$ is transformed not by $\mu_j$ ($j\neq k$) but by $\mu_k$ as
\begin{align}
x_{k;(\ell+1)(N+1)}
&=
\left(\mu_N\circ\mu_{N-1}\circ\cdots\circ\mu_1\right)(x_{k;\ell(N+1)})
\nonumber\\
&=
\mu_k(x_{k;\ell(N+1)})
\nonumber\\
&=
x_{k;\ell(N+1)+k}
\nonumber\\
&=
\frac{y_{k;\ell(N+1)+k-1}x_{k-1;(\ell+1)(N+1)}x_{k+1;\ell(N+1)}+1}
{(y_{k;\ell(N+1)+k-1}\oplus 1)x_{k;\ell(N+1)}},
\label{eq:mutcvprecise}
\end{align}
where we use the equalities
\begin{align*}
x_{k-1;\ell(N+1)+k-1}
&=
\left(\mu_{N+1}\circ\cdots\circ\mu_{k+1}\circ\mu_k\right)(x_{k-1;\ell(N+1)+k-1})
=
x_{k-1;(\ell+1)(N+1)},
\\
x_{k+1;\ell(N+1)+k-1}
&=
\left(\mu_k\circ\cdots\circ\mu_2\circ\mu_1\right)(x_{k+1;\ell(N+1)})
=
x_{k+1;\ell(N+1)}
\end{align*}
derived from the fact that the mutations $\mu_k,\mu_{k+1},\ldots,\mu_{N+1}$ and $\mu_1,\mu_2,\ldots,\mu_k$ do not vary $x_{k-1;\ell(N+1)+k-1}$ and  $x_{k+1;\ell(N+1)+k-1}$, respectively.
Therefore, we need the explicit form of the coefficient $y_{k;\ell(N+1)+k-1}$ ($k=1,2,\ldots,N+1$) to execute the computation.

\begin{proposition}\label{prop:coeffkterm}
For $\ell=nN+s\geq1$, where $0\leq n$ and $1\leq s\leq N$, the coefficient  $y_{k;\ell(N+1)+k-1}$ is explicitly given by
\begin{align*}
y_{k;\ell(N+1)+k-1}
&=
(C_N)^{n(N+1)+s-1}C_{s+k-1}y_{N+1}
\end{align*}
for $k=1,2,\ldots,N+1$.
\end{proposition}

(Proof)\quad
By using theorem \ref{thm:cqgs} and the exchange relation \eqref{eq:ercoeffnew}, we have
\begin{align*}
y_{1;\ell(N+1)}
&=
y_1^\ell
=
(C_N)^{n(N+1)+s-1}C_sy_{N+1},
\\
y_{2;\ell(N+1)+1}
&=
\mu_1(y_{2;\ell(N+1)})
=
y_2^\ell y_1^\ell
=
(C_N)^{n(N+1)+s-1}C_{s+1}y_{N+1}.
\end{align*}
Thus we inductively obtain
\begin{align*}
y_{k;\ell(N+1)+k-1}
&=
\mu_{k-1}(y_{k;\ell(N+1)+k-2})
=
y_{k;\ell(N+1)+k-2}y_{k-1;\ell(N+1)+k-2}
\\
&=
y_k^\ell y_{k-1;\ell(N+1)+k-2}
=
(C_N)^{n(N+1)+s-1}C_{s+k-1}y_{N+1}
\end{align*}
for $k=1,2,\ldots,N+1$.
\qed

\section{Dynamics of cluster variables}
\label{sec:dynamicsofclustervariables}
\subsection{Birational map}
\label{subsec:birationalmap}

Iteration of the consecutive seed mutations $\mu_{N+1}\circ\cdots\circ\mu_2\circ\mu_1$ assigned to the path $\varpi$ of type $A^{(1)}_N$ induces a certain dynamical system governed by a birational map.

Let $\Sigma_m=(\bx_m,\by_m,B_m)$ be the seed assigned to the vertex $t_m$ in the path $\varpi$ of type $A^{(1)}_N$.
For $t\geq0$, we introduce new variables:
\begin{align*}
\bx^t
&:=
\left(x_1^t,x_2^t,\ldots,x_{N+1}^t\right),
\\
x_i^t
&:=
x_{i;t(N+1)}
\quad
(i=1,2,\ldots,N+1).
\end{align*}
Note that $\bx^0$ is the initial cluster $\bx$:
\begin{align*}
\bx^0
=
\left(x_1^0,x_2^0,\ldots,x_{N+1}^0\right)
=
\left(x_1,x_2,\ldots,x_{N+1}\right)
=
\bx
\end{align*}
and $\bx^t$ is the cluster assigned to the vertex $t_{t(N+1)}$ in $\varpi$.

\begin{thm}\label{thm:birationalmapzforz}
The cluster variables assigned to the path $\varpi$ of type $A^{(1)}_N$ are given by using the solutions to the following dynamical system
\begin{align}
\begin{cases}
z_i^{t+1}
=
\DIS\frac{z_{i-1}^{t+1}z_{i+1}^t+1}{z_i^t}
&(i=1,2,\ldots,N+1),
\\
z_{0}^{t+1}
=
z_{N+1}^t,
\\
z_{N+2}^t
=
z_1^{t+1}
\\
\end{cases}
\label{eq:bmA1N}
\end{align}
for $t\geq1$.
\end{thm}

(Proof)\quad
Put $t=nN+s\geq1$ for $0\leq n$ and $1\leq s\leq N$.
By proposition \ref{prop:coeffkterm}, we have
\begin{align*}
y_{i;t(N+1)+i-1}
&=
(C_N)^{n(N+1)+s-1}C_{s+i-1}y_{N+1}
\end{align*}
for $i=1,2,\ldots,N+1$.
Then the exchange relation  \eqref{eq:mutcvprecise} reduces to
\begin{align}
&(C_N)^{n(N+1)+s-1}C_{s+i-1}y_{N+1}x_i^{t+1}x_i^t
=
(C_N)^{n(N+1)+s-1}C_{s+i-1}y_{N+1}x_{i-1}^{t+1}x_{i+1}^t
+
1
\label{eq:xdifferenceequation}
\end{align}
for $i=1,2,\ldots,N+1$, where we assume $x_0^{t+1}=x_{N+1}^t$ and $x_{N+2}^t=x_1^{t+1}$.

Suppose that the variables $z_1^t,z_2^t,\ldots,z_{N+1}^t$ for $t\geq1$ satisfy
\begin{align}
z_i^{t+1}z_i^t
&=
(C_N)^{n(N+1)+s-1}C_{s+i-1}y_{N+1}x_i^{t+1}x_i^t,
\label{eq:xyzrelation1}
\\
z_{i-1}^{t+1}z_{i+1}^t
&=
(C_N)^{n(N+1)+s-1}C_{s+i-1}y_{N+1}x_{i-1}^{t+1}x_{i+1}^t.
\label{eq:xyzrelation2}
\end{align}
Substitute \eqref{eq:xyzrelation1} and \eqref{eq:xyzrelation2} into \eqref{eq:xdifferenceequation}.
Then we see that $z_1^t,z_2^t,\ldots,z_{N+1}^t$ solve \eqref{eq:bmA1N} by setting $z_0^{t+1}=z_{N+1}^t$ and $z_{N+2}^t=z_1^{t+1}$.

We show that \eqref{eq:xyzrelation1} and \eqref{eq:xyzrelation2} are compatible with each other.
First, by \eqref{eq:xyzrelation1}, we have
\begin{align}
x_i^{t+1}
&=
\frac{z_i^{t+1}z_i^t}{(C_N)^{n(N+1)+s-1}C_{s+i-1}y_{N+1}}
\frac{1}{x_i^t}.
\label{eq:xzcompatibility1}
\end{align}
Also, by \eqref{eq:xyzrelation2}, we have
\begin{align}
x_{i-1}^{t+1}
&=
\frac{z_{i-1}^{t+1}z_{i+1}^t}{(C_N)^{n(N+1)+s-1}C_{s+i-1}y_{N+1}}
\frac{1}{x_{i+1}^t}.
\label{eq:xzcompatibility2}
\end{align}
Thus, by successive application of \eqref{eq:xzcompatibility1} and \eqref{eq:xzcompatibility2}, we obtain
\begin{align}
x_i^{t+1}
&=
\frac{z_i^{t+1}z_i^t}{(C_N)^{n(N+1)+s-1}C_{s+i-1}y_{N+1}}
\frac{1}{x_i^t}
\nonumber\\
&=
\frac{z_i^{t+1}z_i^t}{(C_N)^{n(N+1)+s-1}C_{s+i-1}y_{N+1}}
\frac{(C_N)^{n(N+1)+s-2}C_{s+i-1}y_{N+1}}{z_i^tz_{i+2}^{t-1}}
x_{i+2}^{t-1}
\nonumber\\
&=
\frac{1}{C_N}
\frac{z_i^{t+1}}{z_{i+2}^{t-1}}
x_{i+2}^{t-1}.
\label{eq:compatibility}
\end{align}
Similarly, by successive application of \eqref{eq:xzcompatibility2} and \eqref{eq:xzcompatibility1}, we obtain
\begin{align*}
x_i^{t+1}
&=
\frac{z_i^{t+1}z_{i+2}^t}{(C_N)^{n(N+1)+s-1}C_{s+i}y_{N+1}}
\frac{1}{x_{i+2}^t}
\\
&=
\frac{z_i^{t+1}z_{i+2}^t}{(C_N)^{n(N+1)+s-1}C_{s+i}y_{N+1}}
\frac{(C_N)^{n(N+1)+s-2}C_{s+i}y_{N+1}}{z_{i+2}^tz_{i+2}^{t-1}}
x_{i+2}^{t-1}
\\
&=
\frac{1}{C_N}
\frac{z_i^{t+1}}{z_{i+2}^{t-1}}
x_{i+2}^{t-1}.
\end{align*}
This coincides with \eqref{eq:compatibility}.
Therefore, the equations \eqref{eq:xyzrelation1} and \eqref{eq:xyzrelation2} are compatible with each other.
Thus, if the solution $\left(z_1^t,z_2^t,\ldots,z_{N+1}^t\right)$ to the difference equation \eqref{eq:bmA1N} for $t\geq1$ is given then we inductively obtain the solution $\left(x_1^t,x_2^t,\ldots,x_{N+1}^t\right)$ to the difference equation \eqref{eq:xdifferenceequation} for $t\geq1$ by using \eqref{eq:xyzrelation1} or \eqref{eq:xyzrelation2}.
Remark that the cluster variables $x_1^t,x_2^t,\ldots,x_{N+1}^t$ are given by using $x_i^1$ and $z_i^1$ for $i=1,2,\ldots,N+1$.

Now we show that the cluster variables $x_1^t,x_2^t,\ldots,x_{N+1}^t$ can be given by using the initial ones $x_1,x_2,\ldots,x_{N+1}$.
Noting \eqref{eq:xyzrelation1} and \eqref{eq:xyzrelation2}, it is clear that if $z_i^1\propto x_i^1$, that is, $z_i^1$ is proportional to $x_i^1$, for $i=1,2,\ldots,N+1$ then $z_i^t\propto x_i^t$ for $i=1,2,\ldots,N+1$ and $t\geq1$.
Thus if we assume $z_i^1\propto x_i^1$, for $i=1,2,\ldots,N+1$ the cluster variables $x_1^t,x_2^t,\ldots,x_{N+1}^t$ are given by $x_1^1,x_2^1,\ldots,x_{N+1}^1$, and hence by the initial cluster variables $x_1,x_2,\ldots,x_{N+1}$.

Finally, we check that the proportionality $z_i^1\propto x_i^1$, for $i=1,2,\ldots,N+1$ is compatible with  \eqref{eq:xyzrelation1} and \eqref{eq:xyzrelation2}.
Assume $t=1$ ($n=0$ and $s=1$) in \eqref{eq:xyzrelation1} and \eqref{eq:xyzrelation2}:
\begin{align*}
z_i^2z_i^1
&=
C_iy_{N+1}x_i^2x_i^1,
\\
z_{i-1}^2z_{i+1}^1
&=
C_iy_{N+1}x_{i-1}^2x_{i+1}^1.
\end{align*}
Then we have
\begin{align}
z_{i+1}^1
&=
\frac{C_iy_{N+1}x_{i-1}^2x_{i+1}^1}{z_{i-1}^2}
=
\frac{C_iy_{N+1}x_{i-1}^2x_{i+1}^1}{C_{i-1}y_{N+1}x_{i-1}^2x_{i-1}^1}{z_{i-1}^1}
=
\frac{\nu_ix_{i+1}^1}{x_{i-1}^1}{z_{i-1}^1}
\label{eq:znux}
\end{align}
for $i=2,3,\ldots,N$.
Substitute $z_i^1=\alpha_i x_i^1$ ($\alpha_i\in\P$)  into \eqref{eq:znux} we obtain
\begin{align}
\alpha_{i+1}
&=
\alpha_{i-1}\nu_i
=
\alpha_{i-1}y_i^{-1}
\quad
(i=2,3,\ldots,N).
\label{eq:alphay}
\end{align}
For $i=1$, we also obtain
\begin{align}
\alpha_2\alpha_{N+1}=y_1^{-1}
\label{eq:alphayi1}
\end{align}
from \eqref{eq:xyzrelation2}, where we use the boundary conditions $z_0^2=z_{N+1}^1$ and $x_0^2=x_{N+1}^1$.
Thus if $\alpha_2,\alpha_3,\ldots,\alpha_{N+1}$ satisfy \eqref{eq:alphay} and \eqref{eq:alphayi1} then $z_i^1=\alpha_i x_i^1$ holds for $i=1,2,\ldots,N+1$ for any $\alpha_1\in\P$.
\qed

We put
\begin{align*}
\bz^t:=\left(z_1^t,z_2^t,\ldots,z_{N+1}^t\right)
\end{align*}
for $t\geq1$ and denote the birational map $\bz^t\mapsto\bz^{t+1}$ on $\mathcal{F}^{N+1}$ defined by \eqref{eq:bmA1N} by $\varphi$.
We often refer to the birational map $\varphi$ and to the dynamical system \eqref{eq:bmA1N} governed by $\varphi$ as of type $A^{(1)}_N$ as well as the path $\varpi$ from which they are arising.

\subsection{Periodicity}
\label{subsec:pclustervar}
Let us consider the birational map $\varphi$ given by \eqref{eq:bmA1N}.
Denote the Laurent polynomial ring $\ZP\left[(x_1)^\pm,(x_2)^\pm,\ldots,(x_{N+1})^\pm\right]$ generated by the initial cluster variables $x_1,x_2,\ldots,x_{N+1}$ by $\ZP\left[\bx^\pm\right]$.
For simplicity, we denote the polynomial $p(x_1^t,x_2^t,\ldots,x_{N+1}^t)\in\ZP\left[(\bx^t)^\pm\right]$, where $p=p(x_1^1,x_2^1,\ldots,x_{N+1}^1)\in\ZP\left[(\bx^1)^\pm\right]$, by $p^t$ ($p^1=p$).
We use the same notations for the Laurent polynomial ring $\ZP[(\bz^t)^\pm]$ generated by $z_1^t,z_2^t,\ldots,z_{N+1}^t$.
Remark that if we assume $z_i^1\propto x_i^1$ for $i=1,2,\ldots,N+1$ then we have $\ZP[(\bz^t)^\pm]=\ZP[(\bx^t)^\pm]$ for any $t\geq1$ (see the proof of theorem \ref{thm:birationalmapzforz}).

Now we define the Laurent polynomials $\lambda_1,\lambda_2,\ldots,\lambda_N\in\ZP\left[(\bz^1)^\pm\right]$ to be
\begin{align}
\begin{cases}
\DIS \lambda_i
=
\lambda_i(z_1^1,z_2^1,\ldots,z_{N+1}^1)
:=
\frac{z_i^1+z_{i+2}^1}{z_{i+1}^1}
\quad(i=1,2,\ldots,N-1),
\\[10pt]
\DIS \lambda_N
=
\lambda_N(z_1^1,z_2^1,\ldots,z_{N+1}^1)
:=
\frac{z_1^1z_{N}^1+z_2^1z_{N+1}^1+1}{z_1^1z_{N+1}^1}.
\end{cases}
\label{eq:defoflambda}
\end{align}

\begin{lemma}
\label{lem:lambdaLaurent}
If we assume $z_i^1\propto x_i^1$ for $i=1,2,\ldots,N+1$ then we have
\begin{align*}
\lambda_i\in\ZP[\bx^\pm]
\end{align*}
for $i=1,2,\ldots,N$.
\end{lemma}

(Proof)\quad
First note that if we assume $z_i^1=\alpha_i x_i^1$ ($\alpha_i\in\P$) then $\alpha_i$ ($i=2,3,\ldots,N+1$) satisfies \eqref{eq:alphay} and \eqref{eq:alphayi1}.
In addition, remark that the cluster variables $x_i^1$ and $x_i$ satisfy the exchange relation (see \eqref{eq:mutcv} and \eqref{eq:mutcvprecise})
\begin{align}
x_i^1
=
\frac{y_ix_{i-1}^1x_{i+1}+1}{x_i}
\label{eq:x0x1}
\end{align}
for $i=1,2,\ldots,N+1$ and the boundary conditions $x_{0}^1=x_{N+1}$ and $x_{N+2}=x_1^1$, where we use the fact $y_{i;i-1}=y_i$ (see the proof of lemma \ref{lem:y1}).

Assume $i=N+1$.
Then \eqref{eq:x0x1} reduces to
\begin{align*}
x_{N+1}^1x_{N+1}
=
y_{N+1}x_N^1x_{N+2}+1
=
y_{N+1}x_N^1x_1^1+1,
\end{align*}
where we use the boundary condition $x_{N+2}=x_1^1$.
This implies that \eqref{eq:xyzrelation2} reduces to
\begin{align*}
z_N^1z_1^1=y_{N+1}x_N^1z_1^1
\end{align*}
for $i=N+1$ and $t=0$.
Substitution of $z_i^1=\alpha_i x_i^1$ ($i=1,N$) into this equation leads to
\begin{align}
\alpha_1\alpha_N=y_{N+1}.
\label{eq:alphayiN+1}
\end{align}

For $i=1,2,\ldots,N-2$, we compute
\begin{align*}
\lambda_i
&=
\frac{z_i^1+z_{i+2}^1}{z_{i+1}^1}
=
\frac{\alpha_i x_i^1+\alpha_{i+2}x_{i+2}^1}{\alpha_{i+1}x_{i+1}^1}
=
\frac{\alpha_i}{\alpha_{i+1}}
\frac{y_{i+1}x_i^1+x_{i+2}^1}{y_{i+1}x_{i+1}^1}.
\end{align*}
where we use \eqref{eq:alphay}.
Moreover, by using \eqref{eq:x0x1}, we have
\begin{align*}
\lambda_i
&=
\frac{\alpha_i}{\alpha_{i+1}}
\frac{y_{i+1}x_i^1+x_{i+2}^1}{y_{i+1}x_{i+1}^1}
=
\frac{\alpha_i}{\alpha_{i+1}}
\frac{(y_{i+1}x_i^1x_{i+2}+1)+y_{i+2}x_{i+1}^1x_{i+3}}{y_{i+1}x_{i+1}^1x_{i+2}}
\\
&=
\frac{\alpha_i}{\alpha_{i+1}}
\frac{x_{i+1}x_{i+1}^1+y_{i+2}x_{i+1}^1x_{i+3}}{y_{i+1}x_{i+1}^1x_{i+2}}
=
\frac{\alpha_i}{\alpha_{i+1}}
\frac{x_{i+1}+y_{i+2}x_{i+3}}{y_{i+1}x_{i+2}}\in\ZP[\bx^\pm].
\end{align*}

Similarly, we compute
\begin{align*}
\lambda_{N-1}
&=
\frac{\alpha_{N-1}}{\alpha_N}
\frac{x_N+y_{N+1}x_1^1}{y_Nx_{N+1}}
=
\frac{\alpha_{N-1}}{\alpha_N}
\frac{x_1x_N+y_1y_{N+1}x_2x_{N+1}+y_{N+1}}{y_Nx_1x_{N+1}}\in\ZP[\bx^\pm],
\end{align*}
where we use the boundary conditions $x_{0}^1=x_{N+1}$ and $x_{N+2}=x_1^1$.

Noting \eqref{eq:alphayi1} and \eqref{eq:alphayiN+1}, we have
\begin{align*}
\lambda_N
&=
\frac{\alpha_1\alpha_Nx_1^1x_N^1+\alpha_2\alpha_{N+1}x_2^1x_{N+1}^1+1}{\alpha_1\alpha_{N+1}x_1^1x_{N+1}^1}
=
\frac{y_1(y_{N+1}x_1^1x_N^1+1)+x_2^1x_{N+1}^1}{\alpha_1\alpha_{N+1}y_1x_1^1x_{N+1}^1}
\\
&=
\frac{y_1x_{N+1}x_{N+1}^1+x_2^1x_{N+1}^1}{\alpha_1\alpha_{N+1}y_1x_1^1x_{N+1}^1}
=
\frac{y_1x_{N+1}+x_2^1}{\alpha_1\alpha_{N+1}y_1x_1^1}
\\
&=
\frac{(y_1x_2x_{N+1}+1)+y_2x_1^1x_3}{\alpha_1\alpha_{N+1}y_1x_1^1x_2}
=
\frac{x_1x_1^1+y_2x_1^1x_3}{y_1\alpha_1\alpha_{N+1}x_1^1x_2}
=
\frac{x_1+y_2x_3}{\alpha_1\alpha_{N+1}y_1x_2}\in\ZP[\bx^\pm].
\end{align*}
Thus the Laurent polynomials $\lambda_1,\lambda_2,\ldots,\lambda_N\in\ZP[(\bz^1)^\pm]$ are in the Laurent polynomial ring $\ZP[\bx^\pm]$ generated by the initial cluster variables.

Finally, we give the ratio $\alpha_i$ of the variables $z_i^1$ and $x_i^1$, explicitly.
By applying \eqref{eq:alphay} repeatedly, we have
\begin{align*}
\alpha_1\alpha_N
&=
\begin{cases}
\alpha_1\alpha_2(y_3)^{-1}(y_5)^{-1}\cdots (y_{N-1})^{-1}
&(\mbox{$N$ even}),
\\
(\alpha_1)^2(y_2)^{-1}(y_4)^{-1}\cdots (y_{N-1})^{-1}
&(\mbox{$N$ odd}),
\\
\end{cases}
\\
\alpha_2\alpha_{N+1}
&=
\begin{cases}
\alpha_1\alpha_2(y_2)^{-1}(y_4)^{-1}\cdots (y_N)^{-1}
&(\mbox{$N$ even}),
\\
(\alpha_2)^2(y_3)^{-1}(y_5)^{-1}\cdots (y_N)^{-1}
&(\mbox{$N$ odd}).
\\
\end{cases}
\end{align*}
Then \eqref{eq:alphayi1} and \eqref{eq:alphayiN+1} reduce to
\begin{align}
\begin{cases}
\alpha_1\alpha_2
=
(y_1)^{-1}y_2y_4\cdots y_N
&(\mbox{$N$ even}),
\\
(\alpha_2)^2
=
(y_1)^{-1}y_3y_5\cdots y_N
&(\mbox{$N$ odd})
\\
\end{cases}
\label{eq:alphadetermine1}
\end{align}
and
\begin{align}
\begin{cases}
\alpha_1\alpha_2
=
y_3y_5\cdots y_{N+1}
&(\mbox{$N$ even}),
\\
(\alpha_1)^2
=
y_2y_4\cdots y_{N+1}
&(\mbox{$N$ odd}),
\\
\end{cases}
\label{eq:alphadetermine2}
\end{align}
respectively.
Therefore, the ratio $\alpha_i$ ($i=1,2,\ldots,N+1$) is explicitly given by using the initial coefficients $y_1,y_2,\ldots,y_{N+1}$ via \eqref{eq:alphay}, \eqref{eq:alphadetermine1} and \eqref{eq:alphadetermine2}.  
Moreover, from \eqref{eq:alphadetermine1} and \eqref{eq:alphadetermine2}, $y_1,y_2,\ldots,y_{N+1}$ must satisfy
\begin{align}
y_1y_3\cdots y_{N+1}
=
y_2y_4\cdots y_N
\label{eq:assumpevenN}
\end{align}
for even $N$.
\qed

Hereafter, we assume that the initial coefficients $y_1,y_2,\ldots,y_{N+1}$ satisfy \eqref{eq:assumpevenN} for even $N$ unless otherwise stated.

The action of the permutation $\sigma_N\in\mathfrak{S}_N$ (see \eqref{eq:sigmaN}) on $\lambda_i$ is given by
\begin{align*}
\sigma_N\lambda_i
&=
\lambda_{(\sigma_N)^{-1}(i)}
=
\lambda_{i+1}
\end{align*}
for $i=1,2,\ldots,N$.
The Laurent polynomials $\lambda_1,\lambda_2,\ldots,\lambda_N$ have the following periodicity under the evolution by means of the birational map $\varphi$.
\begin{thm}
\label{thm:generator}
For any $t\geq1$ and $i=1,2,\ldots,N$, we have
\begin{align*}
&
\lambda_i^{t+1}
=
\sigma_N \lambda_i^t
=
\lambda_{i+1}^t,
\end{align*}
where the subscript is reduced 
modulo
$N$.
Therefore, every $\lambda_i^t$ has period $N$ on $t$:
\begin{align*}
\lambda_i^{t+N}
=
\lambda_i^t.
\end{align*}
Moreover, we have
\begin{align*}
\lambda_i^t\in\ZP[\bx^\pm]
\end{align*}
for $i=1,2,\ldots,N$ and $t\geq1$.
\end{thm}

(Proof)\quad
For $i=1,2,\ldots,N-2$, by using \eqref{eq:bmA1N}, the Laurent polynomial $\lambda_i^{t+1}$ reduces to
\begin{align*}
\lambda_i^{t+1}
&=
\frac{z_i^{t+1}+z_{i+2}^{t+1}}{z_{i+1}^{t+1}}
=
\frac{z_i^{t+1}{z_{i+2}^t}+z_{i+1}^{t+1}z_{i+3}^t+1}{z_{i+1}^{t+1}{z_{i+2}^t}}
\\
&=
\frac{\left(z_i^{t+1}z_{i+2}^t+1\right){z_{i+1}^t}+\left(z_i^{t+1}z_{i+2}^t+1\right)z_{i+3}^t}{\left(z_i^{t+1}z_{i+2}^t+1\right)z_{i+2}^t}
=
\frac{z_{i+1}^t+z_{i+3}^t}{z_{i+2}^t}
=
\lambda_{i+1}^t.
\end{align*}
For $i=N-1$ and $i=N$, we also compute
\begin{align*}
\lambda_{N-1}^{t+1}
&=
\frac{z_{N-1}^{t+1}+z_{N+1}^{t+1}}{z_N^{t+1}}
=
\frac{z_N^t+z_1^{t+1}}{z_{N+1}^t}
=
\frac{z_1^tz_N^t+z_2^tz_{N+1}^t+1}{z_1^tz_{N+1}^t}
=
\lambda_N^t
\end{align*}
and
\begin{align*}
\lambda_N^{t+1}
&=
\frac{z_1^{t+1}z_N^{t+1}+z_2^{t+1}z_{N+1}^{t+1}+1}{z_1^{t+1}z_{N+1}^{t+1}}
=
\frac{\left(z_N^{t+1}z_1^{t+1}+1\right){z_{N+1}^t}+z_2^{t+1}\left(z_N^{t+1}z_1^{t+1}+1\right)}{z_1^{t+1}\left(z_N^{t+1}z_1^{t+1}+1\right)}
\\
&=
\frac{z_{N+1}^t+z_2^{t+1}}{z_1^{t+1}}
=
\frac{{z_2^t}z_{N+1}^t+z_1^{t+1}z_3^t+1}{z_1^{t+1}{z_2^t}}
\\
&=
\frac{\left(z_{N+1}^tz_2^t+1\right){z_1^t}+\left(z_{N+1}^tz_2^t+1\right)z_3^t}{\left(z_{N+1}^tz_2^t+1\right)z_2^t}
=
\frac{z_1^t+z_3^t}{z_2^t}
=
\lambda_1^t,
\end{align*}
respectively.
Then it is clear that $\lambda_i^t$ ($i=1,2,\ldots,N$) has period $N$ on $t$.
It is also clear from lemma \ref{lem:lambdaLaurent} that we have
\begin{align*}
\lambda_i^t
=
(\sigma_N)^{t-1}\lambda_i
\in\ZP[\bx^\pm]
\end{align*}
for $i=1,2,\ldots,N$ and $t\geq1$.
\qed

We denote the (non-Laurent) polynomial subring $\ZP\left[\lambda_1^t,\lambda_2^t,\ldots,\lambda_N^t\right]$ of the ambient field $\mathcal{F}$ generated by the Laurent polynomials $\lambda_1^t,\lambda_2^t,\ldots,\lambda_N^t$ simply by $\ZP\left[\blambda^t\right]$.

\begin{proposition}\label{prop:PR}
For any $t\geq1$, we have
\begin{align*}
\ZP\left[\blambda^t\right]
=
\ZP\left[\blambda\right]
\subset
\ZP\left[\bx^\pm\right]
\end{align*}
\end{proposition}

(Proof)\quad
Let $f^t$ be a polynomial in $\ZP\left[\blambda^t\right]$.
By using theorem \ref{thm:generator}, we have
\begin{align*}
f^t
&=
f(\lambda_1^t,\lambda_2^t,\ldots,\lambda_N^t)
\\
&=
f((\sigma_N)^{t-1}\lambda_1,(\sigma_N)^{t-1}\lambda_2,\ldots,(\sigma_N)^{t-1}\lambda_N)
=
(\sigma_N)^{t-1}f.
\end{align*}
If $f\in\ZP[\blambda]$ we have $f^t=(\sigma_N)^{t-1}f\in\ZP\left[\blambda\right]$.
Conversely, since $f^t\in\ZP\left[\blambda^t\right]$, $f=(\sigma_N)^{-(t-1)}f^t\in\ZP\left[\blambda^t\right]$ holds.
The inclusion $\ZP\left[\blambda\right]\subset\ZP\left[\bx^\pm\right]$ immediately follows from the fact $\lambda_1,\lambda_2,\ldots,\lambda_N\in\ZP\left[\bx^\pm\right]$.
\qed

Introduce the fundamental symmetric polynomial of degree $n$ generated by the Laurent polynomials $\lambda_1,\lambda_2,\ldots,\lambda_N$ and let it be $q_n$ ($n=1,2,\ldots,N$):
\begin{align*}
q_n
=
q_n(\lambda_1,\lambda_2,\ldots,\lambda_N)
:=\sum_{\substack{I\subset\left\{1,2,\ldots,N\right\}\\|I|=n}}\prod_{i\in I}\lambda_i
\in\ZP\left[\blambda\right].
\end{align*}
We denote $q_n(\lambda_1^t,\lambda_2^t,\ldots,\lambda_N^t)\in\ZP\left[\blambda^t\right]$ simply by $q_n^t$ for $t\geq1$ ($q_n^1=q_n$).
We then have the following corollary to theorem \ref{thm:generator} which states conserved quantities of the dynamical system governed by the map $\varphi$.
\begin{corollary}\label{cor:cqclustervar}
The fundamental symmetric polynomial $q_n^t$ ($n=1,2,\ldots,N$) of degree $n$ generated by the Laurent polynomials $\lambda_1^t,\lambda_2^t,\ldots,\lambda_N^t$ is the conserved quantity of the dynamical system $\bz^{t+1}=\varphi(\bz^t)$ governed by the birational map $\varphi$, that is, we have
\begin{align*}
q_n^t=(\sigma_N)^{t-1}q_n=q_n
\end{align*}
for any $t\geq1$.
\end{corollary}

(Proof)\quad
For any $t\geq1$, since $\sigma_Nq_n=q_n$, it immediately follows
\begin{align*}
q_n^t
&=
q_n\left(\lambda_1^t,\lambda_2^t,\ldots,\lambda_{N}\right)
\\
&=
q_n\left((\sigma_N)^{t-1}\lambda_1,(\sigma_N)^{t-1}\lambda_2,\ldots,(\sigma_N)^{t-1}\lambda_{N}\right)
\\
&=
(\sigma_N)^{t-1}q_n(\lambda_1,\lambda_2,\ldots,\lambda_N)
\\
&=
q_n
\end{align*}
from theorem \ref{thm:generator}.
\qed

In the following subsections, we construct the general solution to the dynamical system $\bz^{t+1}=\varphi(\bz^t)$ governed by the birational map $\varphi$ via linearization of $\varphi$ in terms of the Laurent polynomials $\lambda_1^t,\lambda_2^t,\ldots,\lambda_N^t$.

\subsection{Linearization}
\label{subsec:linearization}
\begin{proposition}\label{prop:linearization}
The quadratic birational map $\varphi:\bz^t\mapsto\bz^{t+1}$ given by \eqref{eq:bmA1N} is equivalent to the non-autonomous linear map $\bz^t\mapsto\bz^{t+1}$ defined by
\begin{align}
\begin{cases}
z_1^{t+1}
=
\lambda_N^tz_{N+1}^t-z_N^t,\\
z_2^{t+1}
=
\lambda_1^tz_1^{t+1}-z_{N+1}^t,\\
z_{i+2}^{t+1}
=
\lambda_{i+1}^{t}z_{i+1}^{t+1}-z_i^{t+1}
&(i=1,2,\ldots,N-1)
\\
\end{cases}
\label{eq:linearmap}
\end{align}
for $t\geq1$.
\end{proposition}

(Proof)\quad
By using the Laurent polynomial $\lambda_N^t$, the difference equation in \eqref{eq:bmA1N} for $i=1$ reduces to
\begin{align*}
z_1^{t+1}
&=
\frac{z_{N+1}^tz_2^t+1}{z_1^t}
=
\frac{\lambda_N^tz_1^tz_{N+1}^t-z_1^tz_N^t}{z_1^t}
=
\lambda_N^tz_{N+1}^t-z_N^t.
\end{align*}
Similarly, by using $\lambda_N^{t+1}$, the one for $i=N+1$ reduces to
\begin{align*}
z_{N+1}^{t+1}
&=
\frac{z_N^{t+1}z_1^{t+1}+1}{z_{N+1}^t}
=
\frac{\lambda_N^{t+1}z_1^{t+1}z_{N+1}^{t+1}-z_2^{t+1}z_{N+1}^{t+1}}{z_{N+1}^t}
=
z_{N+1}^{t+1}\frac{\lambda_N^{t+1}z_1^{t+1}-z_2^{t+1}}{z_{N+1}^t}.
\end{align*}
It follows that we have
\begin{align*}
z_2^{t+1}
&=
\lambda_N^{t+1}z_1^{t+1}-z_{N+1}^t
=
\lambda_1^tz_1^{t+1}-z_{N+1}^t,
\end{align*}
where we use the fact $\lambda_N^{t+1}=\lambda_1^t$.
Moreover, by noticing $\lambda_i^{t+1}=\lambda_{i+1}^t$, the Laurent polynomial $\lambda_i^{t+1}=({z_i^{t+1}+z_{i+2}^{t+1}})/{z_{i+1}^{t+1}}$ can be written as
\begin{align*}
z_{i+2}^{t+1}
&=
\lambda_i^{t+1}z_{i+1}^{t+1}-z_i^{t+1}
=
\lambda_{i+1}^{t}z_{i+1}^{t+1}-z_i^{t+1}
\end{align*}
for $i=1,2,\ldots,N-1$.
\qed

Thus, noting proposition \ref{prop:PR}, we see that the quadratic map $\varphi$ is linearized by using the Laurent polynomials $\lambda_1,\lambda_2,\ldots,\lambda_N$ in $\ZP\left[\bx^{\pm}\right]$.

\begin{remark}\label{rem:linearization}
Hone, Lampe and Kouloukas showed that iteration of the cluster mutations of type $A^{(1)}_N$ has linear degree growth  \cite{HLK19}.
In two dimension, it is well known that linear degree growth of a map iteration leads to linearization of the map \cite{DF01}; however, as far as the authors know, it is not clear whether the map exhibiting linear degree growth is linearizable or not in higher dimensions.
\end{remark}

\begin{remark}
Fordy and Hone constructed the nonlinear recurrence for the T-system equivalent to \eqref{eq:bmA1N} from the same sequence of mutations of cluster variables considered here with all coefficient variables set to 1 \cite{FH14}.
They also gave the quantities $J_n$ which generate the conserved quantities of the nonlinear recurrence and linearize it as well as the Laurent polynomials $\lambda_i$ given by \eqref{eq:defoflambda}.
\end{remark}

\begin{remark}
Keller and Scherotzke showed that all frieze sequences of cluster variables associated with the quiver of type $\widetilde A_{q,q}$ satisfy the linear recurrence relations equivalent to $\lambda_i$ ($i=1,2,\ldots,N-1$) by using the representation-theoretic approach \cite{KS11}.
The quiver of type $\widetilde A_{q,q}$ has the following orientation and labeling
\begin{align*}
\xymatrix{
&&&
\overset{0}{\bigotimes}
& 
\\
\overset{1}{\bigcirc}
\ar@{->}[rrru]
& 
\cdots
\ar@{->}[l]
&
\overset{q-1}{\bigcirc}
\ar@{->}[l]
&
\overset{q}{\bigodot}
\ar@{->}[l]
\ar@{->}[r]
&
\overset{q+1}{\bigcirc}
\ar@{->}[r]
& 
\cdots
\ar@{->}[r]
&
\overset{2q-1}{\bigcirc}.
\ar@{->}[lllu]
}
\end{align*}
\end{remark}

The simultaneous system \eqref{eq:linearmap} of linear equations can be written by using a matrix and vectors as follows
\begin{align}
\left(
\begin{matrix}
1&0&\cdots&\cdots&0\\
-\lambda_1^t&1&\ddots&&\vdots\\
1&-\lambda_2^t&\ddots&\ddots&\vdots\\
\vdots&\ddots&\ddots&1&0\\
0&\cdots&1&-\lambda_N^t&1\\
\end{matrix}
\right)
\left(
\begin{matrix}
z_1^{t+1}\\
z_2^{t+1}\\
\vdots\\
\vdots\\
z_{N+1}^{t+1}\\
\end{matrix}
\right)
=
z_N^{t}
\left(
\begin{matrix}
-1\\
0\\
\vdots\\
\vdots\\
0\\
\end{matrix}
\right)
+
z_{N+1}^{t}
\left(
\begin{matrix}
\lambda_N^t\\
-1\\
0\\
\vdots\\
0\\
\end{matrix}
\right)
\label{eq:sle}
\end{align}

Let us denote the coefficient matrix of \eqref{eq:sle} by $A^t$.
Also denote the vectors $(-1,0,\ldots,0)^T$ and $(\lambda_N^t,-1,0,\ldots,0)^T$ in the right hand side of  \eqref{eq:sle} by $\bb_1$ and $\bb_2$, respectively.
Let the matrix obtained from $A^t$ by replacing its $i$-th column with $\bb_1$ be $A_{i,1}^t$.
Also, let the one by replacing the $i$-th column with $\bb_2$ be $A_{i,2}^t$.
Remark that the determinants $\det A_{i,1}^t$ and $\det A_{i,2}^t$ are in the polynomial ring $\ZP\left[\blambda\right]$ (see proposition \ref{prop:PR}).

Now we consider the following polynomials $\xi_{i,1},\xi_{i,2}\in\ZP\left[\blambda\right]$:
\begin{align*}
\xi_{i,1}
&=
\det A_{i,1}^1
=
-\tilde{a}_{1i},
\\
\xi_{i,2}
&=
\det A_{i,2}^1
=
\lambda_N\tilde{a}_{1i}-\tilde{a}_{2i}
\end{align*}
for $i=1,2,\ldots,N+1$, where $\tilde{a}_{ij}$ is the $(i,j)$-cofactor of the matrix $A^1$.
Note that we have
\begin{align*}
(\sigma_N)^{t-1}\xi_{i,1}
&=
\det A_{i,1}^t,
\\
(\sigma_N)^{t-1}\xi_{i,2}
&=
\det A_{i,2}^t
\end{align*}
for 
any
$t\geq1$ and $(\sigma_N)^{t-1}\xi_{i,1},(\sigma_N)^{t-1}\xi_{i,2}\in\ZP\left[\blambda\right]$ for $i=1,2,\ldots,N+1$.

We, moreover, introduce the $2\times2$ matrix $M$ whose entries are taken from $\ZP\left[\blambda\right]$:
\begin{align*}
M
&:=
\left(
\begin{matrix}
\xi_{N,1}&\xi_{N,2}\\
\xi_{N+1,1}&\xi_{N+1,2}\\
\end{matrix}
\right).
\end{align*}
The action of the permutation $\sigma_N\in\mathfrak{S}_N$ on $M$ is defined to be
\begin{align*}
\sigma_N M
:=
\left(
\begin{matrix}
\sigma_N \xi_{N,1}&\sigma_N \xi_{N,2}\\
\sigma_N \xi_{N+1,1}&\sigma_N \xi_{N+1,2}\\
\end{matrix}
\right).
\end{align*}
The entries of $(\sigma_N)^t M$ are in $\ZP\left[\blambda\right]$ for any $t\geq1$.

\subsection{General solution}
\label{subsec:gsclustervar}
Noting $\det A^t=1$, we solve the system \eqref{eq:sle} of linear equations by using the Cramer formula.
Then we obtain
\begin{align}
z_i^{t+1}
&=
\left((\sigma_N)^{t-1}\xi_{i,1}\right)
z_N^t
+
\left((\sigma_N)^{t-1}\xi_{i,2}\right)
z_{N+1}^t
\label{eq:xsol}
\end{align}
for $i=1,2,\ldots,N+1$.

For $m\geq2$, we put
\begin{align*}
\mathcal{M}_m
:=
\left((\sigma_N)^{m-1}M\right)\left((\sigma_N)^{m-2}M\right)\cdots\left(\sigma_NM\right)M.
\end{align*}
With imposing $i=N$ and $N+1$ to \eqref{eq:xsol}, we obtain
\begin{align}
\left(
\begin{matrix}
z_N^{t+1}\\
z_{N+1}^{t+1}\\
\end{matrix}
\right)
&=
\left((\sigma_N)^{t-1}M\right)
\left(
\begin{matrix}
z_N^{t}\\
z_{N+1}^{t}\\
\end{matrix}
\right)
=
\mathcal{M}_t
\left(
\begin{matrix}
z_N^1\\
z_{N+1}^1\\
\end{matrix}
\right).
\label{eq:xsol2}
\end{align}
Then a theorem which states the general solution to the linear system \eqref{eq:linearmap}, hence to the dynamical system \eqref{eq:bmA1N}, follows.

\begin{thm}\label{thm:GS}
For given $N\geq2$, denote $t\geq1$ by $t=nN+s$, where $0\leq n$ and $1\leq s\leq N$.
Then the general solution to the dynamical system $\bz^{t+1}=\varphi(\bz^t)$ governed by the birational map $\varphi$ is given by
\begin{align}
z_i^t
&=
\left(
(\sigma_N)^{s-1}\xi_{i,1},
(\sigma_N)^{s-1}\xi_{i,2}
\right)
\mathcal{M}_s
\left(\mathcal{M}_N\right)^n
\left(
\begin{matrix}
z_N^1\\
z_{N+1}^1\\
\end{matrix}
\right)
\label{eq:varphisol}
\end{align}
for $i=1,2,\ldots,N+1$.
\end{thm}

(Proof)\quad
Since every $\xi_{i,j}$ has period $N$, $(\sigma_N)^N\xi_{i,j}=\xi_{i,j}$, we have $(\sigma_N)^NM=M$.
Thus \eqref{eq:xsol2} reduces to
\begin{align*}
\left(
\begin{matrix}
z_N^t\\
z_{N+1}^t\\
\end{matrix}
\right)
&=
\mathcal{M}_{nN+s}
\left(
\begin{matrix}
z_N^1\\
z_{N+1}^1\\
\end{matrix}
\right)
=
\mathcal{M}_s
\left(\mathcal{M}_N\right)^n
\left(
\begin{matrix}
z_N^1\\
z_{N+1}^1\\
\end{matrix}
\right).
\end{align*}
Similarly, we obtain
\begin{align*}
z_i^{t+1}
&=
\left((\sigma_N)^{s-1}\xi_{i,1}\right)
z_N^t
+
\left((\sigma_N)^{s-1}\xi_{i,2}\right)
z_{N+1}^t
\end{align*}
from \eqref{eq:xsol}, which leads to \eqref{eq:varphisol}.
\qed

Thus we see that the variables $z_1^t,z_2^t,\ldots,z_{N+1}^t$ have the form
\begin{align*}
z_i^t
&=
h_{i,1}(\lambda_1,\lambda_2,\ldots,\lambda_N)z_N^1
+
h_{i,2}(\lambda_1,\lambda_2,\ldots,\lambda_N)z_{N+1}^1
\end{align*}
for any $t\geq1$, where $h_{i,1}(\lambda_1,\lambda_2,\ldots,\lambda_N),h_{i,2}(\lambda_1,\lambda_2,\ldots,\lambda_N)\in\ZP[\blambda]$.
The assumption $z_i^1\propto x_i^1$ for $i=1,2,\ldots,N+1$ leads to $\ZP[\blambda]\subset\ZP[\bx^\pm]$, hence we have $z_1^t,z_2^t,\ldots,z_{N+1}^t\in\ZP[\bx^\pm]$ for any $t\geq1$.
Since the assumption also leads to  $z_i^t\propto x_i^t$ for $i=1,2,\ldots,N+1$ and $t\geq1$ (see the proof of theorem \ref{thm:birationalmapzforz}), the cluster variables $x_1^t,x_2^t,\ldots,x_{N+1}^t$ assigned to the path of type $A^{(1)}_N$ exhibit the Laurent phenomenon \cite{FZ02}, that is, they are in $\ZP[\bx^\pm]$ for any $t\geq1$.

We see from Corollary \ref{cor:cqclustervar} that the fundamental symmetric polynomials $q_1, q_2,\ldots,q_N$ of the Laurent polynomials $\lambda_1,\lambda_2,\ldots,\lambda_N$ given by \eqref{eq:defoflambda} are the conserved quantities of the dynamical system $\bz^{t+1}=\varphi(\bz^t)$ governed by the nonlinear birational map $\varphi:\mathcal{F}^{N+1}\to\mathcal{F}^{N+1}$ defined by \eqref{eq:bmA1N}.
In Proposition \ref{prop:linearization}, by using the generators $\lambda_1,\lambda_2,\ldots,\lambda_N$ of the conserved quantities, the nonlinear map $\varphi$ is non-autonomously linearized as \eqref{eq:linearmap}.
Through the linearization of $\varphi$, the general solution to the dynamical system governed by $\varphi$ is concretely constructed as \eqref{eq:varphisol} in Theorem \ref{thm:GS}.
Thus, we can say that  the dynamical system governed by $\varphi$ is integrable.
Moreover, by applying Theorem \ref{thm:birationalmapzforz}, the cluster variables in the seed $\Sigma_m=(\bx_m,\by_m,B_m)$ ($m=0,1,2,\ldots$) assigned to the path $\varpi$ of type $A^{(1)}_N$ are given by using the solutions to the dynamical system governed by $\varphi$.
Remark that the arbitrary coefficients in the seed $\Sigma_m$ are also obtained by using Theorem \ref{thm:cqgs}.

It should be note that if we consider the dynamical system $\bz^{t+1}=\varphi(\bz^t)$ in the projective space $\P^{N+1}(\C)$ the generators $\lambda_1,\lambda_2,\ldots,\lambda_N$ of the conserved quantities give the invariant curve of the dynamical system, which is  a union of mutually disjoint quadratic curves each of which is on the 2-dimensional subspace of $\P^{N+1}(\C)$ (see Appendix \ref{sec:Appbirationalmap}).

\section{Concluding remarks}
\label{sec:CONCL}
In the enomous network consisting of infinitely many seeds generated by the mutations $\mu_1,\mu_2,\ldots,\mu_{N+1}$ from the initial seed $\Sigma_0$, we consider the sequence $\Sigma_0,\Sigma_1,\Sigma_2,\ldots$ of seeds whose exchange matrices $B_0,B_1,B_2,\ldots$ respectively correspond to the cluster mutation-periodic quivers $Q_0,Q_1,Q_2,\ldots$ with period 1.
The sequence of seeds is assigned to the path $\varpi$ in the $(N+1)$-regular tree $\T_{N+1}$ by the sub-cluster pattern $\varpi\to\bSigma$.
In the sequence of seeds assigned to the path $\varpi$, every exchange matrix has periodicity with period $N+1$ and has the Cartan counterpart of type $A^{(1)}_N$.
Due to the periodicity of exchange matrices, iteration of the consecutive seed mutations $\mu_{N+1}\circ\cdots\circ\mu_2\circ\mu_1$ induces dynamical systems of the coefficients and of the cluster variables, respectively.
In the dynamics of coefficients, we find that the $N$ monomials $\nu_1^\ell,\nu_2^\ell,\ldots,\nu_N^\ell$ generated by the coefficients $y_1^\ell,y_2^\ell,\ldots,y_{N+1}^\ell$ have the periodicity with period $N$ on $\ell$, and they induce the conserved quantity $C_N$ of the dynamics.
By using these monomials, we obtain the general terms of the coefficients.
Similarly, in the dynamics of cluster variables, we also find the $N$ Laurent polynomials $\lambda_1^t,\lambda_2^t,\ldots \lambda_N^t$ generated by the variables $z_1^t,z_2^t,\ldots,z_{N+1}^t$ associated with the cluster variables $x_1^t,x_2^t,\ldots,x_{N+1}^t$ have the same periodicity on $t$ with $N$.
The Laurent polynomials also induce the conserved quantities of the dynamics.
The dynamics of $z_i^t$ governed by the quadratic birational map $\varphi$ is non-autonomously linearized by virtue of the Laurent polynomials.
Via the linearization of the map $\varphi$, we obtain the general solution to the dynamical system governed by $\varphi$.
Thus the seed $\Sigma_m$ ($m=0,1,2,\ldots$) assigned to the path $\varpi$ is completely solved, that is, the elements of the seed are explicitly given by using the initial ones.
It immediately follows the very well known fact that the cluster variables assigned to the GCM of type $A^{(1)}_N$ via the path $\varpi$ exhibit the Laurent phenomenon.

In the preceding papers \cite{Nobe16, Nobe19}, we considered two kinds of rank 2 seed mutations respectively assigned to the GCMs of types $A^{(1)}_1$ and $A^{(2)}_2$, and showed the integrability of the dynamical systems respectively associated with them.
It followed that the general terms of the cluster variables were concretely constructed by using the conserved quantities.
Moreover, we found that the two dynamical systems are mutually commutative on the conic, which is their common invariant curve, and are linearizable as well as the $A^{(1)}_N$ case which is a generalization of the $A^{(1)}_1$ case.
Therefore, it is natural to expect that the generalized cases of $A^{(2)}_2$, the $A^{(2)}_{2N}$ and $A^{(2)}_{2N+1}$ cases, are integrable and linearizable.
It is also expected that the $A^{(1)}_N$ case is commutative with the $A^{(2)}_{2N}$ and $A^{(2)}_{2N+1}$ cases, respectively.
We will report on this subject in a forthcoming paper.

\begin{acknowledgments}
This work is partially supported by JSPS KAKENHI Grant No. 20K03692.
The authors would like to thank the anonymous referee for careful reading of our manuscript and insightful comments and suggestions.
\end{acknowledgments}

~

{\flushleft\textbf{DATA AVAILABILITY}}

Data sharing is not applicable to this article as no new data were created or analyzed in this study.

\appendix

\section{Birational map $\boldsymbol{\varphi}$ on the projective space}
\label{sec:Appbirationalmap}

\subsection{Homogeneous map}
\label{subsec:homogeneousmap}
Let us consider the birational map $\varphi$ given by \eqref{eq:bmA1N} on the  projective space $\P^{N+1}(\C)$.
It is equivalent to assume the initial point $\bz^0=(z_1^0,z_2^0,\ldots,z_{N+1}^0)$ to be in $\P^{N+1}(\C)$.
We introduce the homogeneous coordinate $(z_1,z_2,\ldots,z_{N+1})\mapsto[W:Z_1:Z_2:\cdots:Z_{N+1}]=[1:z_1,z_2,\ldots,z_{N+1}]$ of the projective space $\P^{N+1}(\C)$.

\begin{proposition}
\label{prop:birationalmaponprojectiveplane}
In the homogeneous coordinate $(z_1,z_2,\ldots,z_{N+1})\mapsto[W:Z_1:Z_2:\cdots:Z_{N+1}]=[1:z_1,z_2,\ldots,z_{N+1}]$ of the projective space $\P^{N+1}(\C)$, the birational map $\varphi$ given by \eqref{eq:bmA1N} reduces to the following homogeneous map of degree $N+3$:
\begin{align}
\begin{cases}
\DIS
W^{t+1}
=
W^tZ_1^t\prod_{j=1}^{N+1}Z_j^t,
\\
\DIS
Z_i^{t+1}
=
Z_1^t\prod_{j=i+1}^{N+1}Z_j^t
\left[
Z_{N+1}^t\prod_{j=2}^{i+1}Z_j^t+(W^t)^2\sum_{j_1,\ldots,j_{i-1}}\prod_{\ell=1}^{i-1}Z_{j_\ell}^t
\right]
\quad (i=1,2,\ldots,N),
\\
\DIS
Z_{N+1}^{t+1}
=
(W^t)^2Z_1^t\prod_{j=1}^NZ_j^t
+
\left[
Z_{N+1}^t\prod_{j=2}^{N+1}Z_j^t+(W^t)^2\sum_{j_1,\ldots,j_{N-1}}\prod_{\ell=1}^{N-1}Z_{j_\ell}^t
\right]
\left[
Z_{N+1}^tZ_2^t+(W^t)^2
\right],
\\
\end{cases}
\label{eq:birationalmaphomogeneouscoordinate}
\end{align}
where if $i=2,3,\ldots,N$ the sum $\DIS\sum_{j_1,\ldots,j_{i-1}}$ in $Z_i^{t+1}$ ranges over the set $\{1,2,\ldots,i+1\}\backslash\{k,k+1\}$ for $k=1,2,\ldots,i$, \textit{i.e.}, 
\begin{align*}
\{j_1,\ldots,j_{i-1}\}
=&
\{3,4,\ldots,i+1\},
\quad
\{1,4,5,\ldots,i+1\},
\quad
\{1,2,5,6,\ldots,i+1\},
\\
&\qquad
\ldots,
\quad
\{1,2,\ldots,i-2,i+1\},
\quad
\{1,2,\ldots,i-1\},
\end{align*}
and if $i=1$ the sum equals $1$.
\end{proposition}

(Proof)\quad
We use induction on $i$.
First, by substitution of the homogeneous coordinate into \eqref{eq:bmA1N} for $i=1$, we have
\begin{align*}
\frac{Z_1^{t+1}}{W^{t+1}}
&=
\frac{Z_{N+1}^tZ_2^t+(W^t)^2}{W^tZ_1^t}.
\end{align*}

For $2\leq i\leq N-1$, we assume that the following holds
\begin{align}
\frac{Z_i^{t+1}}{W^{t+1}}
&=
\frac{\DIS Z_{N+1}^t\prod_{j=2}^{i+1}Z_j^t+(W^t)^2\sum_{j_1,\ldots,j_{i-1}}\prod_{\ell=1}^{i-1}Z_{j_\ell}^t}{\DIS W^t\prod_{j=1}^iZ_j^t}.
\label{eq:inductiononiHC}
\end{align}
Then we have
\begin{align*}
\frac{Z_{i+1}^{t+1}}{W^{t+1}}
&=
\frac{Z_i^{t+1}/W^{t+1}\times Z_{i+2}^t/W^t+1}{Z_{i+1}^t/W^t}
\\
&=
\frac{\DIS \left[Z_{N+1}^t\prod_{j=2}^{i+1}Z_j^t+(W^t)^2\sum_{j_1,\ldots,j_{i-1}}\prod_{\ell=1}^{i-1}Z_{j_\ell}^t\right]Z_{i+2}^t+(W^t)^2\prod_{j=1}^iZ_j^t}
{\DIS \left[W^t\prod_{j=1}^iZ_j^t\right]Z_{i+1}^t}
\\
&=
\frac{\DIS Z_{N+1}^t\prod_{j=2}^{i+2}Z_j^t+(W^t)^2\sum_{j_1,\ldots,j_i}\prod_{\ell=1}^iZ_{j_\ell}^t}{\DIS W^t\prod_{j=1}^{i+1}Z_j^t}.
\end{align*}

Moreover, for $i=N+1$, we have
\begin{align*}
\frac{Z_{N+1}^{t+1}}{W^{t+1}}
&=
\frac{{Z_N^{t+1}}/{W^{t+1}}\times{Z_1^{t+1}}/{W^{t+1}}+1}{{Z_{N+1}^t}/{W^t}}
\\
&=
\frac{\DIS \left[Z_{N+1}^t\prod_{j=2}^{N+1}Z_j^t+(W^t)^2\sum_{j_1,\ldots,j_{N-1}}\prod_{\ell=1}^{N-1}Z_{j_\ell}^t\right]\left[Z_{N+1}^tZ_2^t+(W^t)^2\right]+(W^t)^2Z_1^t\prod_{j=1}^NZ_j^t}
{\DIS W^tZ_1^t\prod_{j=1}^{N+1}Z_j^t}.
\end{align*}
This implies that the homogeneous degree of the map must be $N+3$.
Thus, by respectively multiplying the numerator and the denominator of \eqref{eq:inductiononiHC} by $Z_1^t\prod_{j=i+1}^{N+1}Z_j^t$, we obtain the equations in \eqref{eq:birationalmaphomogeneouscoordinate}.
\qed

\subsection{Invariant curve}
\label{subsec:Invariantcurves}
Consider the dynamical system $\bz^{t+1}=\varphi(\bz^t)$ governed by the map $\varphi$ in the inhomogeneous coordinate
\begin{align}
\P^{N+1}(\C)
&\ni[W:Z_1:Z_2:\cdots:Z_{N+1}]
\nonumber\\
&\mapsto
\left(
\frac{Z_1}{W},
\frac{Z_2}{W},
\ldots,
\frac{Z_{N+1}}{W}
\right)
=
(z_1,z_2,\ldots,z_{N+1})
\in\C^{N+1}
\label{eq:inhomogeneouscoordinate}
\end{align}
for $W\neq0$.
Then the invariant curve of the dynamical system is obtained via the Laurent polynomials $\lambda_1,\lambda_2,\ldots,\lambda_N$ in the following manner.

For $s=0,1,\ldots,N-1$, let the intersection of the hyperplanes
\begin{align*}
\left(\lambda_i(z_1,z_2,\ldots,z_{N+1})-\lambda_i^s=0\right)
\quad
(i=1,2,\ldots,N-1)
\end{align*}
and the hypersurface
\begin{align*}
\left(\lambda_N(z_1,z_2,\ldots,z_{N+1})-\lambda_N^s=0\right)
\end{align*}
in $\C^{N+1}$ be $\gamma^s$:
\begin{align*}
\gamma^s:=\bigcap_{i=1}^N\left(\lambda_i(z_1,z_2,\ldots,z_{N+1})-\lambda_i^s=0\right),
\end{align*}
where
\begin{align*}
\lambda_i^s=\lambda_i(z_1^s,z_2^s,\ldots,z_{N+1}^s)
\end{align*}
for $(z_1^s,z_2^s,\ldots,z_{N+1}^s)\in\C^{N+1}$.
Also let the compactification of $\gamma^s$ in $\P^{N+1}(\C)$ be $\widetilde\gamma^s$.
Denote the union of the compact curves $\widetilde\gamma^s$ by $\Gamma$:
\begin{align*}
\Gamma:=\bigcup_{s=0}^{N-1}\widetilde\gamma^s.
\end{align*}

\begin{proposition}\label{invariantcurve}
The compact curves $\widetilde\gamma^0,\widetilde\gamma^1,\ldots,\widetilde\gamma^{N-1}$ are mutually disjoint quadratic curves each of which is on a 2-dimensional subspace of the projective space $\P^{N+1}(\C)$.
Moreover, the point $\bz^t=(z_1^t,z_2^t,\ldots,z_{N+1}^t)=\varphi^t(\bz^0)$ is on the curve $\widetilde\gamma^s$ for $t\equiv s$ (mod $N$) ($s=0,1,\ldots,N-1$).
Thus the invariant curve of the dynamical system $\bz^{t+1}=\varphi(\bz^t)$ governed by the birational $\varphi$ is the union $\Gamma$ of the quadratic curves $\widetilde\gamma^s$.
\end{proposition}

(Proof)\quad
The intersection of the hyperplanes $\left(\lambda_i(z_1,z_2,\ldots,z_{N+1})-\lambda_i^s=0\right)$ for $i=1,2,\ldots,N-1$ is explicitly given by the solution to the following simultaneous system of linear equations
\begin{align}
\begin{cases}
z_1-\lambda_1^sz_2+z_3=0,
\\
z_2-\lambda_2^sz_3+z_4=0,
\\
\cdots
\\
z_{N-1}-\lambda_{N-1}^sz_N+z_{N+1}=0.
\\
\end{cases}
\label{eq:invariantcurvesubspace}
\end{align}
Since the coefficient matrix of the above system has rank $N-1$, the space of solutions is of dimension 2.
We denote the plane in $\C^{N+1}$ given by \eqref{eq:invariantcurvesubspace} by $p^s$.

The simultaneous system \eqref{eq:invariantcurvesubspace} of equations reduces to
\begin{align*}
\left(
\begin{matrix}
1&-\lambda_1^s&1&\cdots&0\\
0&1&-\lambda_2^s&\ddots&\vdots\\
\vdots&\ddots&\ddots&\ddots&1\\
\vdots&&\ddots&1&-\lambda_{N-2}^s\\
0&\cdots&\cdots&0&1\\
\end{matrix}
\right)
\left(
\begin{matrix}
z_1\\
z_2\\
\vdots\\
\vdots\\
z_{N-1}\\
\end{matrix}
\right)
=
z_N
\left(
\begin{matrix}
0\\
\vdots\\
0\\
-1\\
\lambda_{N-1}^s\\
\end{matrix}
\right)
+
z_{N+1}
\left(
\begin{matrix}
0\\
\vdots\\
\vdots\\
0\\
-1\\
\end{matrix}
\right).
\end{align*}
Then, by solving this, we obtain
\begin{align*}
z_j
&=
\left(\lambda_{N-1}^s\tilde b_{N-1,j}-\tilde b_{N-2,j}\right)z_N-\tilde b_{N-1,j}z_{N+1}
\end{align*}
for $j=1,2,\ldots,N-1$, where $\tilde b_{ij}$ is the $(i,j)$-cofactor of the coefficients matrix.
Thus a point on the plane $p^s$ is given by using two parameters $\alpha^s$ and $\beta^s$:
\begin{align}
\left(
\begin{matrix}
z_1\\
\vdots\\
z_{N-1}\\
z_N\\
z_{N+1}\\
\end{matrix}
\right)
&=
\left(
\begin{matrix}
\lambda_{N-1}^s\tilde b_{N-1,1}-\tilde b_{N-2,1}\\
\vdots\\
\lambda_{N-1}^s\tilde b_{N-1,N-1}-\tilde b_{N-2,N-1}\\
1\\
0\\
\end{matrix}
\right)
\alpha^s
-
\left(
\begin{matrix}
\tilde b_{N-1,1}\\
\vdots\\
\tilde b_{N-1,N-1}\\
0\\
1\\
\end{matrix}
\right)
\beta^s.
\label{eq:palnevectorrep}
\end{align}

The hypersurface $\left(\lambda_N(z_1,z_2,\ldots,z_{N+1})-\lambda_N^s=0\right)$ is the quadratic surface given by
\begin{align}
z_1z_N-\lambda_N^sz_1z_{N+1}+z_2z_{N+1}+1=0.
\label{eq:quadraticsurface}
\end{align}
Therefore, the intersection $\gamma^s$ of the plane $p^s$ and the hypersurface given by \eqref{eq:quadraticsurface} is a quadratic curve on the plane $p^s$.
Substituting $z_1$ and $z_2$ into the equation \eqref{eq:quadraticsurface}, we obtain
\begin{align}
1&+\left(\lambda_{N-1}^s\tilde b_{N-1,1}-\tilde b_{N-2,1}\right)(z_N)^2
+\left(\lambda_N^s\tilde b_{N-1,1}-\tilde b_{N-1,2}\right)(z_{N+1})^2
\nonumber\\
&-\left[\tilde b_{N-1,1}+\tilde b_{N-2,2}-\lambda_N^s\tilde b_{N-2,1}+\lambda_{N-1}^s\left(\lambda_N^s\tilde b_{N-1,1}-\tilde b_{N-1,2}\right)\right]z_Nz_{N+1}=0.
\label{eq:invariantcurvezNzN+1}
\end{align}
Denote the left hand side of \eqref{eq:invariantcurvezNzN+1} by $f(z_N,z_{N+1})$.
Then a point on the quadratic curve $\gamma^s$ is given by \eqref{eq:palnevectorrep} with imposing $f(\alpha^s,\beta^s)=0$.

Substitute \eqref{eq:inhomogeneouscoordinate} into \eqref{eq:invariantcurvezNzN+1}, and remove the denominators by multiplying $W^2$.
If we put $W=0$ then we obtain
\begin{align*}
&\left(\lambda_{N-1}^s\tilde b_{N-1,1}-\tilde b_{N-2,1}\right)(Z_N)^2
+\left(\lambda_N^s\tilde b_{N-1,1}-\tilde b_{N-1,2}\right)(Z_{N+1})^2
\nonumber\\
&\qquad
-\left[\tilde b_{N-1,1}+\tilde b_{N-2,2}-\lambda_N^s\tilde b_{N-2,1}+\lambda_{N-1}^s\left(\lambda_N^s\tilde b_{N-1,1}-\tilde b_{N-1,2}\right)\right]Z_NZ_{N+1}=0.
\end{align*}
Thus we see that the compactification $\widetilde\gamma^s$ of the affine curve $\gamma^s$ has two points at infinity counting with multiplicity.

We assume that the curves $\gamma^{s_1}$ and $\gamma^{s_2}$ ($s_1,s_2\in\{0,1,\ldots,N-1\}$, $s_1\neq s_2$) have a point $P=(\zeta_1,\zeta_2,\ldots,\zeta_{N+1})$ in common.
If $N\geq3$ then \eqref{eq:invariantcurvesubspace} for $s=s_1$ and $s=s_2$ reduce to
\begin{align*}
\begin{cases}
\left(\lambda_1^{s_1}-\lambda_1^{s_2}\right)\zeta_2=0,
\\
\left(\lambda_2^{s_1}-\lambda_2^{s_2}\right)\zeta_3=0,
\\
\cdots
\\
\left(\lambda_{N-1}^{s_1}-\lambda_{N-1}^{s_2}\right)\zeta_N=0.
\\
\end{cases}
\end{align*}
Hence $\zeta_2=\zeta_3=\cdots=\zeta_N=0$.
It immediately follows $\zeta_1=\zeta_{N+1}=0$ from \eqref{eq:invariantcurvesubspace}.
Therefore, the two planes $p^{s_1}$ and $p^{s_2}$ respectively given by \eqref{eq:invariantcurvesubspace} for $s=s_1$ and $s=s_2$ intersect only at the origin $(0,0,\ldots,0)$.
However, the origin never solves \eqref{eq:quadraticsurface}.
Thus there is no intersection point $P$ of the curves $\gamma^{s_1}$ and $\gamma^{s_2}$.
It is clear that the points at infinity of these curves are generically different.
Therefore, the compact curves $\widetilde\gamma^{s_1}$ and $\widetilde\gamma^{s_2}$ do not intersect each other.

\begin{figure}[htbp]
\centering
\includegraphics[scale=1]{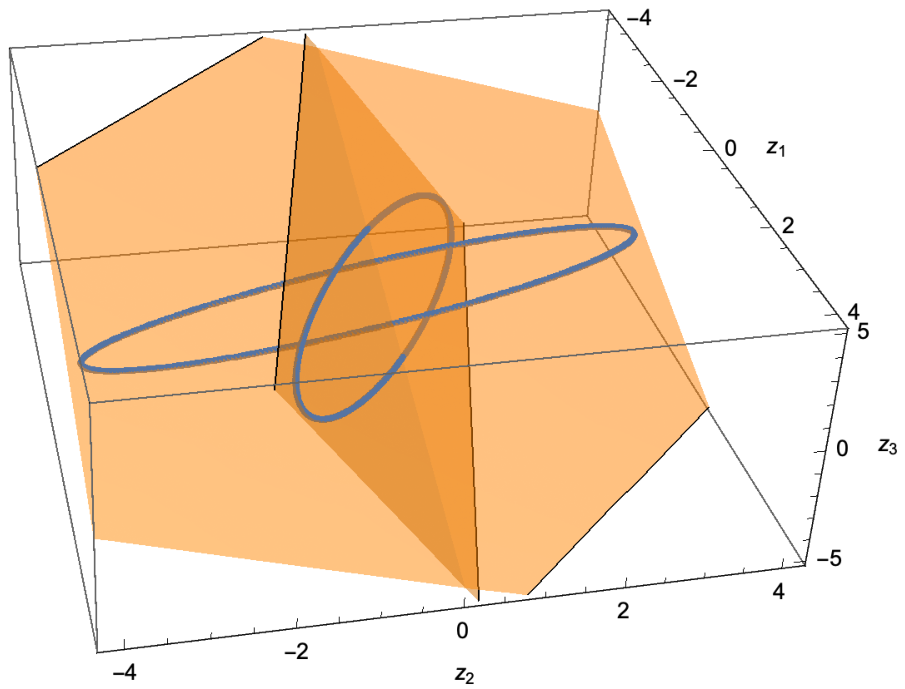}
\caption{The orbit of the dynamical system $\bz^{t+1}=\varphi(\bz^t)$ for $N=2$ in $\R^3$ with imposing the initial values $(z_1,z_2,z_3)=(4,-2,4)\in\R^3$. The invariant curve $\Gamma$ consists of the 2 disjoint quadratic curves $\widetilde\gamma^0$ and $\widetilde\gamma^1$ whose affine parts are on the planes $\left(z_1+4z_2+z_3=0\right)$ and $\left(16z_1+15z_2+16z_3=0\right)$, respectively.
These two planes meet at the line $\left(z_1+z_3=0\right)\cap\left(z_2=0\right)$.}
\label{fig:Orbits}
\end{figure}

If $N=2$ we have $\zeta_2=0$ and $\zeta_1+\zeta_3=0$ from \eqref{eq:invariantcurvesubspace}.
Thus the two planes $p^0$ and $p^1$, respectively given by $\lambda_1=(z_1+z_3)/z_2=\lambda_1^0$ and $\lambda_1=\lambda_1^1$, meet at the line $\left(z_1+z_3=0\right)\cap\left(z_2=0\right)$.
Substituting $\zeta_2=0$ into \eqref{eq:quadraticsurface} for $s=0$ and $s=1$, we obtain
\begin{align*}
(\lambda_2^0-\lambda_2^1)\zeta_1\zeta_3=0.
\end{align*}
It follows that $\zeta_1=\zeta_3=0$.
Therefore, there is no intersection point $P$ of the curves $\gamma^0$ and $\gamma^1$, and hence is of the compact curves $\widetilde\gamma^0$ and $\widetilde\gamma^1$ (see figure \ref{fig:Orbits}).

Since $\lambda_i^t$ has the periodicity (see theorem \ref{thm:generator})
\begin{align*}
\lambda_i^{t+N}
=
\lambda_i^t,
\end{align*}
the point $\bz^t=(z_1^t,z_2^t,\ldots,z_{N+1}^t)=\varphi^t(\bz^0)$ is on the curve $\widetilde\gamma^s$ for $t\equiv s$ (mod $N$).
\qed

The $(N+1)$-dimensional dynamical system governed by the birational map $\varphi$ has $N$ conserved quantities $q_1,q_2,\ldots,q_N$ which are functionally independent (see corollary \ref{cor:cqclustervar}).
Therefore, the dynamical system is integrable in the sense of Liouville.
The invariant curve is the union of the $N$ disjoint quadratic curves each of which is on the 2-dimensional subspace of $\P^{N+1}(\C)$.

\nocite{*}
\bibliography{Nobe190806r}

\end{document}